\documentclass[letterpaper]{article}
\usepackage{amsmath}
\usepackage{amsfonts}
\newcommand{\block}[4]{#1 \otimes #2 \otimes #3 \otimes #4}
\newcommand{\void}[1]{}
\begin{document}

\begin{center}
\mathversion{bold} {\bf Minimal Unitary Realizations of Exceptional
U-duality Groups and Their Subgroups  as Quasiconformal Groups
\footnote{Work supported in
   part by the National Science Foundation under grant number
   PHY-0245337.} }
\bigskip\bigskip
\mathversion{normal}

{ M.~G\"{u}naydin and O.~Pavlyk } \\
{ Penn State University,\\Physics Department,\\
  University Park, PA 16802} \smallskip\\ {\small E-mails:
  murat@phys.psu.edu; pavlyk@phys.psu.edu} \bigskip

\end{center}

\begin{abstract}
We study the minimal unitary representations of noncompact
exceptional groups that arise as U-duality groups in extended
supergravity theories. First we give the unitary realizations of the
exceptional group $\mathrm{E}_{8(-24)}$ in  $\mathrm{SU}^{\ast}(8)$
as well as $\mathrm{SU}(6,2)$ covariant bases. $\mathrm{E}_{8(-24)}$
has $\mathrm{E}_7\times \mathrm{SU}(2)$ as its maximal compact
subgroup and is the U-duality group of the exceptional supergravity
theory in $d=3$. For the corresponding U-duality group
$\mathrm{E}_{8(8)}$ of the maximal supergravity theory the  minimal
realization was given in hep-th/0109005. The minimal unitary
realizations of all the lower rank noncompact exceptional groups can
be obtained by truncation of those of $\mathrm{E}_{8(-24)}$ and
$\mathrm{E}_{8(8)}$. By further truncation one can obtain the
minimal unitary realizations of all the groups of the "Magic
Triangle". We give explicitly the minimal unitary realizations of
the exceptional subgroups of $\mathrm{E}_{8(-24)}$ as well as other
physically interesting subgroups. These minimal unitary realizations
correspond , in general, to the quantization of their geometric
actions as quasi-conformal groups as defined in hep-th/0008063.

\end{abstract}

\section{Introduction}
The concept of a minimal unitary representation of a non-compact group $G$ was first introduced
 by A. Joseph \cite{Joseph}. It is defined as a unitary representation on a Hilbert space of functions
depending on the minimal number of coordinates for a given non-compact group. By introducing position operators
corresponding to these coordinates and the momenta conjugate to them, one obtains the minimal realization by
expressing the generators of the Lie algebra $\mathfrak{g}$ of $G$ in terms of these canonical operators.
Joseph's main motivation was to give a general mathematical framework for spectrum generating symmetry algebras
that were studied by physicists earlier \cite{spectgenalg}.

The minimal realization of complex forms of classical Lie algebras and of $\mathfrak{g}_{2(2)}$ were given by
Joseph \cite{Joseph,Joseph2} in a Cartan-Weyl basis. Over the last two decades there has  been an ever increasing
interest by the mathematicians on the minimal unitary representations of noncompact groups. For a review and the
references on earlier work on the subject we refer the reader to the review lectures of Jian-Shu Li
\cite{li2000}.

 More recently, minimal unitary representations were studied by Pioline, Kazhdan and Waldron \cite{PKW} and by
G\"unaydin, Koepsell and Nicolai \cite{GKN:2} with the goal of applying them to M-theory. The work of KPW was
motivated the idea that the theta series of $\mathrm{E}_{8(8)}$ and its subgroups may describe the quantum supermembrane
in various dimensions \cite{PW}. On the other hand the work of GKN was motivated by the idea that the spectra of
M-theory in various dimensions must fall into unitary representations of its U-duality group in the respective
dimensions. Realization of the minimal unitary representation of $\mathrm{E}_{8(8)}$ and its subalgebras given in  \cite{GKN:2}
is based on an earlier work \cite{GKN:1} of the authors on geometric realization of $\mathrm{E}_{8(8)}$ as a quasi-conformal
group acting on a 57-dimensional space with a quartic norm form.

The groups $\mathrm{E}_{6(6)}$, $\mathrm{E}_{7(7)}$ and $\mathrm{E}_{8(8)}$ arise as  U-duality groups of
maximal supergravity theories
obtained from the 11 dimensional supergravity \cite{11dsugra} by toroidal compactification to $d=5,4$ and $d=3$
dimensions, respectively. For M-theory that has the 11 dimensional supergravity as its low energy effective
theory in a strongly coupled phase one expects only the discrete subgroups $\mathrm{E}_{6(6)}(\mathbb{Z})$,
$\mathrm{E}_{7(7)}(\mathbb{Z})$ and $\mathrm{E}_{8(8)}(\mathbb{Z})$ to be symmetries of the full
nonperturbative theory. Hence we
expect the spectra of M-theory to fall into unitary representations of these discrete subgroups of U-duality
groups in the respective compactifications.

As was shown in \cite{GKN:1} $\mathrm{E}_{8(8)}$ has a natural action as a quasi-conformal group in the charge-entropy
space of BPS black hole solutions in $N=8$ supergravity in $d=4$ and hence can be interpreted as its spectrum
generating symmetry group. The formula relating the entropy of a four dimensional BPS black hole solution to its
charges defines a generalized light-cone in the charge-entropy space which is left invariant by the
quasiconformal group action of $\mathrm{E}_{8(8)}$.

In addition to $\mathrm{E}_{8(8)}$ the only other non-compact real form of
$\mathrm{E}_8$ is $\mathrm{E}_{8(-24)}$ whose maximal compact
subgroup is $\mathrm{E}_7 \, \otimes \,\mathrm{SU}(2)$.
It is the U-duality group of the exceptional $\mathcal{N}=4$ supergravity in
$d=3$, which  can be obtained, by dimensional reduction, from the exceptional $\mathcal{N}=2$ Maxwell-Einstein
supergravity theory (MESGT) in $d=5$ describing the coupling of 26 vector multiplets to pure $\mathcal{N}=2$
supergravity \cite{gst}. In five dimensions U-duality group of the exceptional MESGT is $\mathrm{E}_{6(-26)}$ with
maximal compact subgroup  $\mathrm{F}_4$ and  in $d=4$ it is  $\mathrm{E}_{7(-25)}$  with maximal
compact subgroup $\mathrm{E}_6 \otimes
\mathrm{U}(1)$.

In this paper we extend the results of \cite{GKN:2} on the minimal unitary representations of $\mathrm{E}_{8(8)}$ to the
other noncompact real form  $\mathrm{E}_{8(-24)}$ and study their  truncations to other exceptional quasi-conformal
subalgebras as well as other classical subalgebras. We should note that, just like  $\mathrm{E}_{8(8)}$, the group
$\mathrm{E}_{8(-24)}$ can be realized as a quasi-conformal group acting on a 57-dimensional space. For the exceptional
$\mathcal{N}=2$ $d=4$ supergravity theory this 57-dimensional space is
 the charge-entropy space and $\mathrm{E}_{8(-24)}$ acts as a spectrum generating quasiconformal symmetry that leaves the generalized
 light-cones invariant.

The plan of the paper is as follows. In section 2 we write down the
Lie algebra of the exceptional group $\mathrm{E}_{8(-24)}$ and then
give its minimal unitary realization in a $\mathrm{SU}^{\ast}(8)$ as
well as $\mathrm{SU}(6,2)$ covariant bases. The
$\mathrm{SU}^{\ast}(8)$ basis is the coordinate basis (Schr\"odinger
picture) while the $\mathrm{SU}(6,2)$ basis is the oscillator
realization. They correspond to the $\mathrm{SL}(8,\mathbb{R})$ and
$\mathrm{SU}(8)$ bases of the maximally split exceptional group
$\mathrm{E}_{8(8)}$ with maximal compact subgroup $SO(16)$  given in
\cite{GKN:2}. In section 3 we give the minimal unitary realizations
of $\mathrm{E}_{7(-5)}$ with maximal compact subgroup
$\mathrm{SO}(12) \times \mathrm{SU}(2)$, $\mathrm{E}_{6(2)}$ with
maximal compact subgroup $\mathrm{SU}(6)\times \mathrm{SU}(2)$,
$\mathrm{E}_{6(-14)}$ with the maximal compact subgroup
$\mathrm{SO}(10)\times \mathrm{U}(1)$,  $\mathrm{F}_{4(4)} $ with
maximal compact subgroup $\mathrm{USp}(6)\times \mathrm{USp}(2)$ and
$\mathrm{SO}(4,4)$ by consistent truncation of the minimal unitary
realization of $\mathrm{E}_{8(-24)}$. In section 4 we study a
different chain  of truncations and give the minimal unitary
realization of $\mathrm{E}_{7(-25)}$ with maximal compact subgroup
$E_6\times \mathrm{U}(1)$ and of $\mathrm{SO}(2p,2)$ for
$p=2,3,4,5$. We conclude with a discussion of our results and future
directions. In appendix A we give the explicit transformations for
going from the $\mathrm{SU}^{\ast}(8)$ basis to the
$\mathrm{SU}(6,2)$ of the minimal realization of
$\mathrm{E}_{8(-24)}$ and in appendix B we give the minimal unitary
realization of $\mathrm{E}_{8(8)}$ in a $\mathrm{SU}^{\ast}(8)$
covariant basis.

\section{Minimal Unitary Representation of $\mathrm{E}_{8(-24)}$}
The Lie algebra $\mathfrak{e}_{8(-24)}$ of $\mathrm{E}_{8(-24)}$ admits a
5-grading with respect its subalgebra $\mathfrak{e}_{7(-25)}
\oplus \mathfrak{so}(1,1)$ determined by the  generator $\Delta$
of a dilatation subgroup $SO(1,1)$
\begin{equation}
 \mathfrak{e}_{8(-24)} \mspace{5mu} = \mspace{5mu}
\begin{array}{ccccccccc}
   \mathfrak{g}^{-2} & \oplus & \mathfrak{g}^{-1} & \oplus & \mathfrak{g}^{0} & \oplus &
   \mathfrak{g}^{+1} & \oplus & \mathfrak{g}^{+2} \\
   \mathbf{1} & \oplus & \mathbf{56} & \oplus & \left( \mathbf{133} \oplus \mathbf{1} \right) & \oplus &
   \mathbf{56} & \oplus & \mathbf{1}
\end{array}
\label{eq:decomp}
\end{equation}
such that $\mathfrak{g}^{\pm 2}$ generators together with $\Delta$ form an $\mathfrak{sl}(2, \mathbb{R})$
subalgebra.

To construct the minimal unitary representation of $\mathfrak{e}_{8(-24)}$ we find it convenient to work in a
basis covariant with respect to $\mathfrak{su}^\ast(8)$ subalgebra of $\mathfrak{e}_{7(-25)}$\footnote{ The
\label{ftnt:aboutSU} $\mathfrak{su}^\ast(8)$-covariant basis of $\mathfrak{e}_{7(-25)}$ is the analog of
$\mathfrak{sl}(8, \mathbb{R})$ basis of $\mathfrak{e}_{7(7)}$ \cite{GKN:1}}. In the $\mathfrak{su}^\ast(8)$ basis
the generators of $\mathfrak{e}_{7(-25)}$ can be labelled as follows
\begin{equation}
    \mathbf{133} = \mathbf{63} \oplus \mathbf{70} = {J^A}_B \oplus J^{ABCD}
\end{equation}
where ${J^A}_B$  denote the generators of $\mathfrak{su}^\ast(8)$ and
 $J^{ABCD}$ is completely antisymmetric in its indices $A,B, \ldots
=1,2,\ldots ,8$.
 They satisfy the  commutation relations
\begin{equation}
\begin{split}
  \left[ {J^A}_B, {J^C}_D \right] & = {\delta^C}_B {J^A}_D - {\delta^A}_D {J^C}_B \\
  \left[ {J^A}_B, J^{CDEF} \right] &= - 4 {\delta^{[C}}_B J^{DEF]A} - \frac{1}{2} {\delta^A}_B J^{CDEF} \\
  \left[ J^{ABCD}, J^{EFGH} \right] &= - \frac{1}{36} \epsilon^{ABCDK[EFG} {J^{H]}}_K
\end{split}
\label{eq:e7}
\end{equation}
and the following reality conditions
\begin{equation}
\begin{split}
   \left( {J^A}_B \right)^\dagger & =  {J_A}^B =  \Omega_{AC} \Omega^{BD} {J^C}_D \\
   \left( J^{ABCD} \right)^\dagger &= - J_{ABCD} = - \Omega_{AE} \Omega_{BF} \Omega_{CG} \Omega_{DH} J^{EFGH} 
\end{split}
\label{eq:realityConditions}
\end{equation}
where $\Omega$ is a symplectic matrix such that $\Omega_{AB}= - \Omega_{BA} = \left( \Omega^{BA} \right)^\ast$,
$\Omega_{AB} \Omega^{BC} = {\delta^C}_A$. The quadratic Casimir operator of $\mathrm{E}_{7(-25)}$ in the basis
\eqref{eq:e7} is given by
\begin{equation}
\begin{split}
 \mathcal{C}_2 & =  \frac{1}{6} {J^A}_B {J^B}_A - \frac{1}{24} \epsilon_{ABCDEFGH} J^{ABCD} J^{EFGH} \\
               & =  \frac{1}{6} {J^A}_B {J^B}_A - J^{ABCD} (\epsilon J)_{ABCD}
\end{split}
 \label{eq:e7C2}
\end{equation}
where  $ \left(\epsilon J\right)_{ABCD} = \frac{1}{4!}\epsilon_{ABCDEFGH} J^{EFGH}$.

The fundamental representation  $\mathbf{56}$ of $\mathfrak{e}_{7(-25)}$ decomposes as $\mathbf{28} \oplus
\Tilde{\mathbf{28}}$ under its $\mathfrak{su}^\ast(8)$ subalgebra, where $\mathbf{28}$ ($X^{AB}$) and
$\Tilde{\mathbf{28}}$ ($\Tilde{X}_{AB}$) are anti-symmetric tensors satisfying the following reality condition
\begin{equation}
   \left( X^{AB} \right)^\dagger = X_{AB} = \Omega_{AC} \Omega_{BD} X^{CD} \,, \quad
   \left( \Tilde{X}_{AB} \right)^\dagger =  \Tilde{X}^{AB} = \Omega^{AC} \Omega^{BD} \Tilde{X}_{CD} \,.
\end{equation}
Under the action of $\mathfrak{e}_{7(-25)}$ they transform as
\begin{equation}\begin{split}
   \delta X^{AB} &= {\Sigma^A}_C X^{CB} + {\Sigma^B}_C X^{AC} - \Sigma^{ABCD} \Tilde{X}_{CD} \\
   \delta \Tilde{X}_{CD} &= - {\Sigma^A}_C \Tilde{X}_{AD} - {\Sigma^A}_D \Tilde{X}_{CA} + \Sigma_{CDAB} X^{AB}
\end{split}
\label{eq:act56}
\end{equation}
where ${\Sigma^A}_C$ and $\Sigma^{ABCD} = - \left( \Sigma_{ABCD}\right)^\dagger$
denote parameters of $\mathrm{SU}^\ast\left(8\right)$ transformation
 and those of  the coset generators $\mathrm{E}_{7(-25)}/\mathrm{SU}^\ast\left(8\right)$ , respectively.

\subsection{Exceptional Lie Algebra $\mathfrak{e}_{8(-24)}$ \label{sec:beg}}
Note that $\mathbf{56}$ is a real representation of
$\mathfrak{e}_{7(-25)}$ just as $\mathbf{28}$ and
$\Tilde{\mathbf{28}}$ are real representations of
$\mathfrak{su}^\ast(8)$. Thus in $\mathfrak{su}^\ast(8)$ covariant
basis we can label generators belonging to grade -1 space as
$E^{AB}$ and $\Tilde{E}_{AB}$ and grade +1 space as $F^{AB}$ and
$\Tilde{F}_{AB}$. The 5-graded decomposition of
$\mathfrak{e}_{8(-24)}$ in $\mathfrak{su}^\ast(8)$ basis takes the
form
\begin{equation}
 \mathfrak{e}_{8(-24)} =  E \oplus \left\{ E^{AB} \,, \Tilde{E}_{CD} \right\} \oplus \left\{ {J^A}_B \,, J^{ABCD} \,;
                    \Delta \right\} \oplus
  \left\{ F^{AB} \,, \Tilde{F}_{CD} \right\} \oplus F
  \tag{\ref{eq:decomp}}
\end{equation}
The grading is defined by the generator $\Delta$ of $SO(1,1)$
\begin{equation}
\begin{split}
  \left[ \Delta, E \right] = -2 E \,, \mspace{6mu} \qquad \qquad \left[\Delta, F \right] = +2 F \mspace{41mu}  \\
  \left[ \Delta, E^{AB} \right] = - E^{AB} \,, \qquad \left[ \Delta, F^{AB} \right] = + F^{AB} \\
  \left[ \Delta, \Tilde{E}_{CD} \right] = - \Tilde{E}_{CD} \,, \qquad
  \left[ \Delta, \Tilde{F}_{CD} \right] = +  \Tilde{F}_{CD}
\end{split}
\end{equation}
Positive and negative generators  form two separate maximal
Heisenberg subalgebras  with commutation relations
\begin{equation}
  \left[ E^{AB} ,\, \Tilde{E}_{CD} \right] = 2 \, \delta^{AB}_{CD} \, E \qquad
  \left[ E,\, E^{AB} \right] = 0 \qquad
  \left[ E,\, \Tilde{E}_{AB} \right] = 0
\end{equation}
and
\begin{equation}
  \left[ F^{AB} \,, \Tilde{F}_{CD} \right] = 2 \, \delta^{AB}_{CD} \, F \qquad
  \left[ F,\, F^{AB} \right] = 0 \qquad
  \left[ F,\, \Tilde{F}_{AB} \right] = 0\,.
\end{equation}
However these two Heisenberg subalgebras do not commute with each other (see eqs.~\eqref{eq:closingintog0} below).
Generators of $\mathfrak{g}^{\pm 2}$ are invariant under $\mathfrak{e}_{7(-25)}$
\begin{equation}
   \left[ {J^A}_B, \, F \right] = 0 \quad \left[ J^{ABCD}, \, F \right] = 0 \quad
   \left[ {J^A}_B, \, E \right] = 0 \quad \left[ J^{ABCD}, \, E \right] = 0
\end{equation}
while generators of $\mathfrak{g}^{\pm 1}$ transform under $\mathfrak{su}^\ast(8)$ as follows
\begin{equation}
\begin{split}
  \left[ {J^A}_B \,, E^{CD} \right] & = {\delta^C}_B E^{AD} + {\delta^D}_B E^{CA} - \frac{1}{4} {\delta^A}_B E^{CD}  \\
  \left[ {J^A}_B \,, F^{CD} \right] & = {\delta^C}_B F^{AD} + {\delta^D}_B F^{CA} - \frac{1}{4} {\delta^A}_B F^{CD}  \\
  \left[ {J^A}_B \,, \Tilde{E}_{CD} \right] & = -{\delta^A}_C \Tilde{E}_{BD} - {\delta^A}_D \Tilde{E}_{CB} +
                                                \frac{1}{4} {\delta^A}_B \Tilde{E}_{CD} \\
  \left[ {J^A}_B \,, \Tilde{F}_{CD} \right] & = -{\delta^A}_C \Tilde{F}_{BD} - {\delta^A}_D \Tilde{F}_{CB} +
                                                \frac{1}{4} {\delta^A}_B \Tilde{F}_{CD}
\end{split}
\end{equation}
The remaining commutation relations read as follows
\begin{equation*}
\begin{split}
   \left[ J^{ABCD} \,, \Tilde{E}_{EF} \right] =  \delta^{[AB}_{EF} E^{CD]} \,, \quad
   \left[ J^{ABCD} \,, E^{EF} \right] = - \frac{1}{24} \epsilon^{ABCDEFGH} \Tilde{E}_{GH} \, \\
   \left[ J^{ABCD} \,, \Tilde{F}_{EF} \right] =  \delta^{[AB}_{EF} F^{CD]} \,, \quad
   \left[ J^{ABCD} \,, F^{EF} \right] = - \frac{1}{24} \epsilon^{ABCDEFGH} \Tilde{F}_{GH} \,
\end{split}
\end{equation*}
\begin{equation}\label{eq:closingintog0}
\begin{split}
 \left[ E^{AB} \,, F^{CD} \right] = -12 \, J^{ABCD} \,, \qquad
\left[ \Tilde{E}_{AB} \,, F^{CD} \right] = 4 \,{\delta^{[C}}_{[A} {J^{D]}}_{B]} + \delta^{CD}_{AB} \Delta \\
 \left[ \Tilde{E}_{AB} \,, \Tilde{F}_{CD} \right] = - 12\, {(\epsilon J)}_{ABCD} \,, \mspace{-19mu}\qquad
 \left[ E^{AB} \,, \Tilde{F}_{CD} \right] = 4 \,{\delta^{[A}}_{[C} {J^{B]}}_{D]} - \delta^{AB}_{CD} \Delta
\end{split}
\end{equation}
\begin{equation*}
\begin{array}{lr}
\begin{aligned}
   \left[ E, F^{AB} \right] = - E^{AB} \,, \quad \left[ E, \Tilde{F}_{AB} \right] = - \Tilde{E}_{AB} \\
   \left[ F, E^{AB} \right] = + F^{AB} \,, \quad \left[ F, \Tilde{E}_{AB} \right] = + \Tilde{F}_{AB}
\end{aligned}
 & \mspace{15mu}   \left[ E, F \right] =  \Delta
\end{array}
\end{equation*}
Reality properties for generators belonging to grade $\pm 1$ and $\pm 2$ are as follows
\begin{equation}
\begin{split}
\begin{aligned}
  &\left(F^{AB}\right)^\dagger = - \Omega_{AC} \Omega_{BD} F^{CD} \,, \quad
  & \left( \Tilde{F}_{AB} \right)^\dagger = - \Omega^{AC} \Omega^{BD} \Tilde{F}_{CD} \,,
\\
   &\left(E^{AB}\right)^\dagger = - \Omega_{AC} \Omega_{BD} E^{CD} \,, \quad
  & \left( \Tilde{E}_{AB} \right)^\dagger = - \Omega^{AC} \Omega^{BD} \Tilde{E}_{CD} \,,
\end{aligned} \\
  E^\dagger = -E \,, \mspace{176mu}  F^\dagger = -F \,, \mspace{99mu}
\end{split}
\label{eq:reality}
\end{equation}
The quadratic Casimir operator of the above Lie algebra is given by
\begin{equation}
\begin{split}
   \mathcal{C}_2 \left( \mathfrak{e}_{8(-24)} \right) &=
     \frac{1}{6} {J^A}_B {J^B}_A - J^{ABCD} (\epsilon J)_{ABCD} \\
     & + \frac{1}{12} \Delta^2 - \frac{1}{6} \left( F E + E F \right) \\
     & - \frac{1}{12} \left( \Tilde{E}_{AB} F^{AB} +  F^{AB} \Tilde{E}_{AB} -
                              \Tilde{F}_{AB} E^{AB} -  E^{AB} \Tilde{F}_{AB}\right)
\end{split}
\label{eq:e8qC}
\end{equation}
In order to make manifest  the fact that the above Lie algebra is
of the real form  $\mathfrak{e}_{8(-24)}$ with the maximal compact
subalgebra $\mathfrak{e}_{7} \oplus \mathfrak{su}(2)$ let us write
down the compact and noncompact generators explicitly. Under the
maximal compact subalgebra $\mathfrak{usp}(8)$ of
$\mathfrak{su}^\ast(8)$ we have the following decompositions of
the adjoint and fundamental representations of
$\mathfrak{e}_{8(-24)}$
\begin{equation*}
\begin{split}
  \mathbf{133} &= \mathbf{63} \oplus \mathbf{70} = \left( \mathbf{36} \oplus \mathbf{27} \right) \oplus
      \left( \mathbf{1} \oplus \mathbf{27} \oplus \mathbf{42} \right) \\
      \mathbf{56} &= \mathbf{28} \oplus \mathbf{\tilde{28}}= ( \mathbf{1} \oplus \mathbf{27} ) \oplus ( \mathbf{1} \oplus
      \mathbf{27} )
\end{split}
\end{equation*}
where $\mathbf{27}$ and $\mathbf{42}$ correspond to symplectic traceless antisymmetric 2-tensor and 4-tensor
of $\mathfrak{usp}(8)$  respectively.\footnote{
The group $\mathrm{SU}^\ast(8)$ is defined as a subgroup of $\mathrm{SL}(8, \mathbb{C})$ generated by
elements $U \in \mathrm{SL}(8,
\mathbb{C})$ such that $U^\dagger U = 1$ and $U \Omega = U^\ast \Omega$. $U^\ast$ is obtained from $U$ by
component-wise complex conjugation.}
Note that the generators in the representations $\mathbf{1} \oplus \mathbf{36} \oplus \mathbf{42}$ of
$\mathfrak{usp}(8)$ in the decomposition of the adjoint representation of $\mathfrak{e}_{7(-25)}$
 form the maximal compact subalgebra  $\mathfrak{e}_6 \oplus
\mathfrak{u}(1) $  of $ \mathfrak{e}_{7(-25)}$.

Denoting the generators ($T$)  transforming covariantly under the $\mathfrak{usp}(8)$ subalgebra of
$\mathfrak{su}^\ast(8)$ with a check ($\Check{T}$) we find that the generators in $\mathbf{36} \oplus
\mathbf{27}$ are given by $\Check{G}^{(\pm)}_{AB} = \Omega_{AC} {J^C}_B \pm \Omega_{BC} {J^C}_A$, while
generators coming from the decomposition of $\mathbf{70}$ with respect to $\mathfrak{usp}(8)$ are given by
\begin{equation*}
  \Check{J}^{AB} = J^{ABCD} \Omega_{CD} + \frac{1}{8} \Omega^{AB} \Check{J} \,.
\end{equation*}
\begin{equation*}
\Check{J}^{ABCD} := J^{ABCD} + \frac{3}{2} \Omega^{[AB} \Omega_{EF} J^{CD]EF}  +
        \frac{1}{8} \Omega^{[AB} \Omega^{CD]} \Check{J}
\end{equation*}
\begin{equation*}
\Check{J} := \Omega_{EF} \Omega_{GH} J^{EFGH}
\end{equation*}
Thus we find that
\begin{equation}
  J^{ABCD}\left(\epsilon J\right)_{ABCD} = \Check{J}^{ABCD}\Check{J}_{ABCD} - \frac{3}{2} \, \Check{J}^{AB} \Check{J}_{AB} + \frac{1}{16} \Check{J}^2
\end{equation}
The decomposition of $\mathbf{56}$ of $ \mathfrak{e}_{7(-25)}$ into $\mathfrak{usp}(8)$ irreducible components
leads to the following generators that transform in the $\mathbf{27}$ of $\mathfrak{usp}(8)$:
\begin{equation}
\begin{split}
   \Check{C}_{AB}^{\pm} &= \Tilde{E}_{AB} + F_{AB} \pm \left( \Tilde{F}_{AB} - E_{AB} \right) \\
                          &+ \frac{1}{8} \Omega_{AB}  \Omega^{CD} \left[\Tilde{E}_{CD} + F_{CD}
                             \pm \left( \Tilde{F}_{CD} - E_{CD} \right) \right]  \\
   \Check{N}_{AB}^{\pm} & =\Tilde{F}_{AB} + E_{AB} \pm \left( \Tilde{E}_{AB} - F_{AB} \right)\\
                          &+ \frac{1}{8} \Omega_{AB} \Omega^{CD} \left[\Tilde{F}_{CD} + E_{CD}
                               \pm \left( \Tilde{E}_{CD} - F_{CD} \right) \right]
\end{split}
\end{equation}
and to the following singlets of $\mathfrak{usp}(8)$:
\begin{eqnarray}
\Check{C}^{\pm} =\Omega^{CD} [\Tilde{E}_{CD} + F_{CD} \pm \left( \Tilde{F}_{CD} - E_{CD} \right) ] \nonumber \\
\Check{N}^{\pm} =\Omega^{CD} [\Tilde{F}_{CD} + E_{CD} \pm \left( \Tilde{E}_{CD} - F_{CD} \right) ] \nonumber
\end{eqnarray}
Then the following 133  operators
\begin{equation}
   \Check{G}^{(+)}_{AB} \,, \quad \Check{J}^{ABCD} \,, \quad \Check{J} + 2 \left( E + F\right) \,, \quad
    \Check{C}_{AB}^{\pm}
\end{equation}
generate the compact $E_7$ subgroup  and the operators  $\Check{C}^{\pm}$ and $2(E+F)-3 \Check{J}$  generate the compact
$SU(2)$ subgroup. The remaining  112 generators are non-compact:
\begin{equation}
   G^{(-)}_{AB} \,, \quad \Check{J}^{AB} \,, \quad \Delta\,, \quad F-E \,, \quad
   \Check{N}_{AB}^{\pm} \, , \quad \Check{N}^{\pm}  .
\end{equation}

\subsection{The Minimal Unitary Realization of $\mathfrak{e}_{8(-24)}$  in $\mathfrak{su}^{\ast}(8)$ Basis}
It was noted earlier that elements of the subspace
$\mathfrak{g}^{-2} \oplus \mathfrak{g}^{-1} \subset
\mathfrak{e}_{8(-24)}$ form an Heisenberg algebra with 28
``coordinates'' and 28 ``momenta'' with the  generator of $\mathfrak{g}^{-2}$
acting as its central charge. As it was done for
$\mathfrak{e}_{8(8)}$ \cite{GKN:1} we shall realize these
Heisenberg algebra generators using canonically conjugate position
($X^{AB}$) and momentum ($P_{AB}$) operators:
\begin{equation}
   \left[ X^{AB} \,, P_{CD} \right] = i \, \delta_{CD}^{AB} \,.
   \label{eq:XPsympl}
\end{equation}
satisfying the following reality properties
\begin{equation}
   \left( X^{AB} \right)^\dagger = X_{AB} = \Omega_{AC} \Omega_{BD} X^{CD} \,,\quad
   \left( {P}_{AB} \right)^\dagger = {P}^{AB} = \Omega^{AC} \Omega^{BD} {P}_{CD}
\end{equation}
The commutation relations \eqref{eq:XPsympl} can also be rewritten in more $\mathfrak{usp}(8)$
covariant fashion
\begin{equation}
   \left[ X_{AB} \,, P_{CD} \right] = \frac{i}{2} \left( \Omega_{AC} \Omega_{BD} -
    \Omega_{BC} \Omega_{AD} \right) . \tag{\ref{eq:XPsympl}$^\prime$}
\end{equation}
The generators of $\mathfrak{g}^{-1} \oplus \mathfrak{g}^{-2}$ subalgebra are then realized as
\begin{equation}
  E^{AB} = -i y \,X^{AB} \qquad
  \Tilde{E}_{AB} = -i y\, P_{AB} \qquad E= -\frac{i}{2} \, y^2
  \label{eq:sustar8negativegrade}
\end{equation}
where $y$ is an extra coordinate related to central charge. In order to be able to realize
$\mathfrak{g}^{+1} \oplus \mathfrak{g}^{+2}$ generators we need to introduce a  momentum
operator $p$ conjugate to $y$:
\begin{equation}
   \left[ y \,, p \right]  =i \,.
\end{equation}
The grade zero $\mathfrak{g}^0$ generators, realized linearly on operators $X^{AB}$ and $P_{AB}$,
take on the form
\begin{equation} \label{eq:e7-25generators}
\begin{split}
  {J^A}_B &= -2 i X^{AC} P_{CB} -  \frac{i}{4} {\delta^A}_B  X^{CD}  P_{CD} \\
  J^{ABCD} &= - \frac{i}{2} X^{[AB} X^{CD]} - \frac{i}{48} \epsilon^{ABCDEFGH} P_{EF} P_{GH}
\end{split}
\end{equation}
The dilatation generator $\Delta$ that defines the grading is simply
\begin{equation}
   \Delta = - \frac{i}{2} \left( py + yp \right).
\end{equation}
Since $\mathfrak{g}^{-1}$ generators are linear and $\mathfrak{g}^{0}$ generators are quadratic polynomials
in $X$ and $P$ we expect $\mathfrak{g}^{+1}$ generators to be cubic. Furthermore,
$\mathfrak{g}^{+1} = \left[ \mathfrak{g}^{+2}, \mathfrak{g}^{-1} \right]$ suggests that $F$ must be
a quartic polynomial in $X$ and $P$. Since it is an $\mathfrak{e}_{7(-25)}$ singlet, this quartic must be
the quartic invariant of $\mathfrak{e}_{7(-25)}$. Indeed we find
\begin{equation}
\begin{split}
  & F = \frac{1}{2 i} p^2 + \frac{2}{i y^2} I_4 \left( X\,, P \right) \\
  & F^{AB} = i p X^{AB} + \frac{2}{y} \left[ X^{AB} \,, I_4 \left( X, P \right) \right] \\
  &  \Tilde{F}_{AB} = i p P_{AB} + \frac{2}{y} \left[ P_{AB} \,, I_4 \left( X, P \right) \right] .
\end{split}
\label{eq:sustar8grade1}
\end{equation}
The quartic invariant $I_4$ coincides with quadratic Casimir of $\mathfrak{e}_{7(-25)}$ modulo an additive constant:
\begin{equation}
\begin{split}
  I_4  \left( X \,, P \right) &= \mathcal{C}_2 \left(\mathfrak{e}_{7(-25)} \right) + \frac{323}{16} = \frac{547}{16} +\\
      & -\frac{1}{2} \left( X^{AB} P_{BC}
    X^{CD} P_{DA} + P_{AB} X^{BC} P_{CD} X^{DA} \right) \\
    &+ \frac{1}{8} \left(
        X^{AB} P_{AB} X^{CD} P_{CD} +
    P_{AB} X^{AB} P_{CD} X^{CD} \right) \\
   & + \frac{1}{96} \epsilon^{ABCDMNKL}
     P_{AB} P_{CD} P_{MN} P_{KL} \\
   & + \frac{1}{96} \epsilon_{ABCDMNKL}
            X^{AB} X^{CD} X^{MN} X^{KL}
\end{split}
\label{eq:I4}
\end{equation}

The quadratic Casimir of $\mathfrak{e}_{8(-24)}$ \eqref{eq:e8qC} evaluated in the above realization reduces to a
c-number as required by the irreducibility. In order to demonstrate that we decompose the quadratic Casimir
\eqref{eq:e8qC} into three $\mathfrak{e}_{7(-25)}$-invariant pieces
\begin{equation*}
   \mathcal{C}_2 \left( \mathfrak{e}_8 \right) = \mathcal{C}_2 \left(  \mathfrak{e}_7 \right)
                      + \mathcal{C}_2 \left( \mathfrak{sl} \left( 2 ,  \mathbb{R} \right) \right) +
                      \mathcal{C}'
\end{equation*}
according to the first, second and the third lines of \eqref{eq:e8qC} respectively. From \eqref{eq:I4} we find that
\begin{equation*}
  \mathcal{C}_2 \left(  \mathfrak{e}_7 \right) =  I_4 - \frac{323}{16}\,.
\end{equation*}
Using definitions of $\Delta, E, F$ we obtain
\begin{equation*}
   \mathcal{C}_2 \left( \mathfrak{sl} \left( 2 ,  \mathbb{R} \right) \right) = \frac{1}{3} I_4 - \frac{1}{16}
\end{equation*}
Using definitions for $\mathfrak{g}^{-1} \oplus \mathfrak{g}^{+1}$ generators we find
\begin{equation*}
\begin{split}
  3 \mathcal{C}' &= 7 - 28 I_4 - i X^{AB} I_4 P_{AB} + i P_{AB} I_4 X^{AB} \\
               &= 7 - 28 I_4 + \left( 32 I_4 - \frac{265}{4} \right) = 4 I_4 - \frac{237}{4}
\end{split}
\end{equation*}
and therefore
\begin{equation} \label{eq:e8-24c2Casimir}
   \mathcal{C}_2 \left( \mathfrak{e}_8 \right) = - 40 \,.
\end{equation}

Since $E_8$ does not have any invariant tensors in 58 dimensions (corresponding to 29 position and 29 momentum
operators) all higher Casimir operators of $\mathfrak{e}_{8(-24)}$ in the above realization must also reduce to
c-numbers as was argued for the case of $\mathfrak{e}_{8(8)}$ in \cite{GKN:2}. By integrating the above Lie
algebra  one obtains the minimal unitary representation of the group ${E}_{8(-24)}$ over the Hilbert space of
square integrable complex  functions in 29 variables.

\subsection{The Minimal Unitary Realization of $\mathfrak{e}_{8(-24)}$ in $\mathfrak{su}(6,2)$ Basis}
 Analysis above was done in $\mathfrak{su}^\ast(8)$ covariant basis (see footnote on the page \pageref{ftnt:aboutSU}).
Since covariant operators $X^{AB}$ and $P_{AB}$ are position and momenta we refer to this basis as the
Schr\"odinger picture. One can consider an oscillator basis where the natural operators are 28 creation and 28
annihilation operators constructed out of $X$ and $P$'s. Being complex, we expect them to transform as
$\mathbf{28} \oplus \overline{\mathbf{28}}$ of some non-compact version of $\mathfrak{su}(8)$ within
 $\mathfrak{e}_{7(-25)} \subset \mathfrak{g}^0$. This algebra turns out to be $\mathfrak{su}(6,2)$ and the
 creation and annihilation operators are given as follows
\begin{equation}
\begin{split}
     Z^{ab} &= \frac{1}{4} \, {\Gamma^{ab}}_{CD}  \left( X^{CD} - i P_{CD} \right) \\
     \Tilde{Z}^{ab} &= \frac{1}{4} \, {\Gamma^{ab}}_{CD}  \left( X^{CD} + i P_{CD} \right)
\end{split}
\end{equation}
where transformation coefficient ${\Gamma^{ab}}_{CD}$ are related to gamma-matrices of
\begin{equation*}
   \mathfrak{so}(6,2) \simeq \mathfrak{so}^\ast(8) \simeq \mathfrak{su}^\ast(8) \cap \mathfrak{su}(6,2)
\end{equation*}
as spelled out in appendix A. Operators $Z$ and $\Tilde{Z}$ satisfy
\begin{equation}
  \left[ \Tilde{Z}^{ab} \,, Z^{cd} \right] = \frac{1}{2} \left( \eta^{ca}\eta^{db} - \eta^{cb}\eta^{da} \right) \,.
\end{equation}
with the following reality conditions
\begin{equation}
    \left(Z^{ab}\right)^\dagger = \Tilde{Z}^{ab} = \eta^{ac}\eta^{bd} \Tilde{Z}_{cd}
\end{equation}
where $\eta = \mathop\mathrm{Diag} \left( +, +, +, +, +, +, -, -\right)$ is used to raise and lower indexes.
Generators of $\mathfrak{e}_{7(-25)}$ in this basis take the following form
\begin{equation}
\begin{split}
   {J^{a}}_b &= 2 {Z}^{ac} \Tilde{Z}_{bc} - \frac{1}{4} {\delta^a}_b {Z}^{cd} \Tilde{Z}_{cd} \\
   J^{abcd} &= \frac{1}{2} {Z}^{[ab} {Z}^{cd]} -
                  \frac{1}{48} \epsilon^{abcdefgh} \Tilde{Z}_{ef} \Tilde{Z}_{gh}
\end{split}
\end{equation}
with hermiticity conditions
\begin{equation}
   \left( {J^{a}}_b \right)^\dagger = \eta^{ad} \eta_{bc} {J^c}_d \qquad
   \left( J^{abcd} \right)^\dagger = - \frac{1}{24} {\epsilon^{abcd}}_{efgh} J^{efgh}
\end{equation}
Their commutation relations are
\begin{equation}
\begin{split}
  \left[ {J^a}_b \,, {J^c}_d \right] &= {\delta^c}_b {J^a}_d - {\delta^d}_a {J^c}_b \\
  \left[ {J^a}_b \,, J^{cdef} \right] &= -4 {\delta^{[c}}_b J^{def]a} - \frac{1}{2} {\delta^a}_b J^{cdef} \\
  \left[ J^{abcd} \,, J^{efgh} \right] &=  \frac{1}{36} \epsilon^{abcdp[efg} {J^{h]}}_p
\end{split}
\end{equation}
which have the same form as $\mathfrak{su}^\ast\left(8\right)$ covariant
eqs.~\eqref{eq:e7}. Quadratic Casimir in this basis reads as
\begin{equation}
\begin{split}
   \mathcal{C}_2 \left( \mathfrak{e}_7 \right) &=  \frac{1}{6} {J^a}_b {J^b}_a + J^{abcd} \left( \epsilon J \right)_{abcd} =
         I_4 \left( Z \,, \Tilde{Z} \right) - \frac{323}{16} = \\
     &  = \frac{1}{2} \left( \Tilde{Z}_{ab} Z^{bc} \Tilde{Z}_{cd} Z^{da} +
                     Z^{ab} \Tilde{Z}_{bc} Z^{cd} \Tilde{Z}_{da} \right)\\
       &  - \frac{1}{8} \left( \Tilde{Z}_{ab} Z^{ab} \Tilde{Z}_{cd} Z^{cd} +
                  Z^{ab} \Tilde{Z}_{ab}  Z^{cd} \Tilde{Z}_{cd}  \right) + 14 \\
      &  \mspace{10mu} + \frac{1}{96} \epsilon_{abcdefgh} Z^{ab} Z^{cd} Z^{ef} Z^{gh}
        + \frac{1}{96} \epsilon^{abcdefgh} \Tilde{Z}_{ab}
                          \Tilde{Z}_{cd} \Tilde{Z}_{ef} \Tilde{Z}_{gh}
\end{split}
\end{equation}
Negative grade generators of $\mathfrak{e}_{8(-24)}$ are then simply
\begin{equation}
  E = \frac{1}{2} y^2  \qquad E^{ab} = y Z^{ab} \qquad \qquad \Tilde{E}_{ab} = y \Tilde{Z}_{ab}
  \label{eq:su62negativegrade}
\end{equation}
Generators in $\mathfrak{g}^{+1}$ can be inferred commuting $\mathfrak{g}^{+2}$ generator
\begin{equation}
   F = \frac{1}{2} p^2 + 2 y^{-2} I_4
\end{equation}
with generators in $\mathfrak{g}^{-1}$
\begin{equation}
\begin{split}
   F^{ab} & = i \left[ E^{ab}\,, F \right] =  - p Z^{ab} + 2 i y^{-1} \left[ Z^{ab} \,, I_4 \right] \\
   \Tilde{F}_{ab} & = i \left[ \Tilde{E}_{ab}\,, F \right] =
          - p \Tilde{Z}_{ab} + 2 i y^{-1} \left[\Tilde{Z}_{ab} \,, I_4 \right]
\end{split}
\label{eq:su62grade1}
\end{equation}
or more explicitly
\begin{equation}
\begin{split}
   F^{ab} = &- p Z^{ab} - \frac{i}{12} y^{-1} \epsilon^{abcdefgh} \Tilde{Z}_{cd} \Tilde{Z}_{ef} \Tilde{Z}_{gh} \\
       & + 4 i y^{-1} Z^{c[a}\Tilde{Z}_{cd} Z^{b]d} + \frac{i}{2} \, y^{-1} \left( Z^{ab} \Tilde{Z}_{cd} Z^{cd} +
       Z^{cd}  \Tilde{Z}_{cd} Z^{ab}\right) \\
  F_{ab} = & -p \Tilde{Z}_{ab} + \frac{i}{12} y^{-1} \epsilon_{abcdefgh} Z^{cd} Z^{ef} Z^{gh} \\
        & - 4 i y^{-1} \Tilde{Z}_{c[a} Z^{cd} Z_{b]d} - \frac{i}{2}\, y^{-1} \left( \Tilde{Z}_{ab} Z^{cd} \Tilde{Z}_{cd}
        + \Tilde{Z}_{cd} Z^{cd} \Tilde{Z}_{ab}  \right)
\end{split}
\end{equation}
We see that commutation relations in this basis closely follow those in $\mathfrak{su}^\ast(8)$ basis,
with modified reality conditions (cf. \eqref{eq:su62negativegrade} and \eqref{eq:sustar8negativegrade}
as well as \eqref{eq:su62grade1} with \eqref{eq:sustar8grade1}).
The $SU(6,2)$ covariant commutation relations  follow closely those given in section \ref{sec:beg}
\begin{equation}
\begin{array}{c}
   \left[ E, F \right] = - \Delta \\[6pt]
\begin{aligned}
  \left[ \Delta,\, F \right] &= 2 F \cr
  \left[ \Delta,\, F^{ab} \right] &= F^{ab} \cr
  \left[ \Delta,\, \Tilde{F}_{ab} \right] &= \Tilde{F}_{ab}
\end{aligned}
 \quad
\begin{aligned}
  \left[ \Delta,\, E \right] &= - 2 E \cr
  \left[ \Delta,\, E^{ab} \right] &= - E^{ab} \cr
  \left[ \Delta,\, \Tilde{E}_{ab} \right] &= - \Tilde{E}_{ab}
\end{aligned}
\end{array}
 \quad
\begin{aligned}
   \left[ E \,, F^{ab} \right] &= - i E^{ab} \cr
   \left[ E \,, \Tilde{F}_{ab} \right] &= - i \Tilde{E}_{ab} \cr
   \left[ F \,, E^{ab} \right] &=  i F^{ab} \cr
   \left[ F \,, \Tilde{E}_{ab} \right] &=  i \Tilde{F}_{ab}
\end{aligned}
\end{equation}
\begin{equation}
\begin{aligned}
  \left[ E,\, E^{ab} \right] &= 0 \cr
  \left[ F,\, F^{ab} \right] &= 0 \cr
\end{aligned}
\quad
\begin{aligned}
  \left[ E,\, \Tilde{E}_{ab} \right] &= 0 \cr
  \left[ F,\, \Tilde{F}_{ab} \right] &= 0
\end{aligned}
\quad
\begin{aligned}
  \left[ \Tilde{E}_{ab} \,, {E}^{cd} \right] &= 2 \, \delta^{cd}_{ab} \, E \cr
   \left[ \Tilde{F}_{ab} \,, {F}^{cd} \right] &= 2 \, \delta^{cd}_{ab} \, F
\end{aligned}
\end{equation}
\begin{equation}
\begin{aligned}
   \left[ E^{ab} \,, F^{cd} \right] &= - 12 i J^{abcd} \cr
   \left[ \Tilde{E}_{ab} ,\, \Tilde{F}_{cd} \right] &=  12 i \left( \epsilon J\right)_{abcd}
\end{aligned}
\quad
\begin{aligned}
   \left[ \Tilde{E}_{ab} ,\, F^{cd} \right] &= - 4 i {\delta^{[c}}_{[a} {J^{d]}}_{b]}
          - i \delta^{cd}_{ab} \Delta  \cr
   \left[ E^{ab} ,\, \Tilde{F}_{cd} \right] &= - 4 i {\delta^{[a}}_{[c} {J^{b]}}_{d]}
          + i \delta^{ab}_{cd} \Delta  \cr
\end{aligned}
\end{equation}
\begin{equation*}
  \begin{aligned}
  \left[ {J^a}_b \,, E^{cd} \right] & = {\delta^c}_b E^{ad} + {\delta^d}_b E^{ca} - \frac{1}{4} {\delta^a}_b E^{cd}  \cr
  \left[  \Tilde{E}_{cd}  \,, {J^a}_b \right] & = {\delta^a}_c \Tilde{E}_{bd} + {\delta^a}_d \Tilde{E}_{cb} -
                                                \frac{1}{4} {\delta^a}_b \Tilde{E}_{cd}
  \end{aligned}
\,
\begin{aligned}
   \left[ J^{abcd} \,, \Tilde{E}_{ef} \right] &=  \delta^{[ab}_{ef} E^{cd]} \cr
   \left[ J^{abcd} \,, E^{ef} \right] &=  \frac{-1}{24} \epsilon^{abcdefgh} \Tilde{E}_{gh}
\end{aligned}
\end{equation*}
The quadratic Casimir of $\mathfrak{e}_{8(-24)}$ in this basis reads as follows
\begin{equation}
\begin{split}
  \mathcal{C}_2 &= \frac{1}{6} \,{J^a}_b {J^b}_a + J^{abcd} \left( \epsilon J\right)_{abcd}
    + \frac{1}{6} \left( E F + F E +\frac{1}{2} \,\Delta^2 \right) \\
   & -
   \frac{i}{12} \left(  \Tilde{F}_{ab} E^{ab} +  E^{ab} \Tilde{F}_{ab}- \Tilde{E}_{ab} F^{ab} - F^{ab} \Tilde{E}_{ab}
   \right)
\end{split}
\end{equation}
and reduces to the same c-number as \eqref{eq:e8-24c2Casimir}.

\section{Truncations of the minimal unitary realization of $\mathfrak{e}_{8(-24)}$}

Since our realization of $\mathfrak{e}_{8(-24)}$ is non-linear,
not every subalgebra of $\mathfrak{e}_{8(-24)}$ can be obtained by
a consistent truncation. We consider consistent truncations to subalgebras
that are quasi-conformal. Since quasi-conformal algebras admit  a 5-grading
\begin{equation*}
 \mathfrak{g}=  \mathfrak{g}^{-2} \oplus \mathfrak{g}^{-1} \oplus \mathfrak{g}^{0} \oplus \mathfrak{g}^{+1} \oplus
  \mathfrak{g}^{+2}
\end{equation*}
with  $\mathfrak{g}^{\pm 2}$ being  one-dimensional, they  have an
$\mathfrak{sl}(2,   \mathbb{R})$  subalgebra generated by elements
of  $\mathfrak{g}^{\pm 2}$  and the generator $\Delta$ that
determines 5-grading.  However, the quartic invariant
$\mathcal{I}_4$ will now be that of a subalgebra $\mathfrak{g}^0$
of the linearly realized $\mathfrak{e}_{7(-25)}$ within
$\mathfrak{e}_{8(-24)}$. Furthermore, this subalgebra must act on
the grade $\pm 1$ subspaces via a \emph{symplectic
representation}.

Hence, the problem is reduced to enumeration of subalgebras of
linearly realized $\mathfrak{e}_{7(-25)}$ admitting a
non-degenerate quartic invariant on the symplectic representation.
Before giving the explicit truncations below  we shall first
indicate a  partial web of consistent truncations as
quasiconformal subalgebras.

Firstly, we can truncate $\mathfrak{e}_{8(-24)}$ down to either $\mathfrak{e}_{7(5)}$
or $\mathfrak{e}_{7(-25)}$, by keeping singlets of either $\mathfrak{su}(2)$ or
$\mathfrak{su}(1,1)$ within $\mathfrak{su}(6,2) \subset \mathfrak{e}_{7(-25)}$ correspondingly.
Further truncations of $\mathfrak{e}_{7(-25)}$ to rank 6 quasi-conformal algebras
can lead to either $\mathfrak{so}(10,2)$ or $\mathfrak{e}_{6(-14)}$, while truncations
of $\mathfrak{e}_{7(5)}$ lead to either $\mathfrak{e}_{6(-14)}$ or $\mathfrak{e}_{6(2)}$:
\begin{equation}
  \mathfrak{e}_{8(-24)} \begin{aligned}\nearrow\cr\searrow\end{aligned}
    \begin{aligned}
           &                       &  & \mathfrak{so}(10,2) \rightarrow  \cr
           &\mathfrak{e}_{7(-25)}  & \begin{aligned}\nearrow\cr\searrow\end{aligned}  &                       \cr
           &                       &  & \mathfrak{e}_{6(-14)} \rightarrow \cr
           &\mspace{15mu}\mathfrak{e}_{7(5)}  &   \begin{aligned}\nearrow\cr\searrow\end{aligned}  &                        \cr
           &                       &  & \mathfrak{e}_{6(2)} \rightarrow  \cr
    \end{aligned}
    \mspace{-40mu}
    \begin{aligned}
         & \mspace{26mu} \mathfrak{so}(6,2) \, \rightarrow\,  \mathfrak{so}(4,2) \cr
         & \\[2pt]
         & \cr
         & \mspace{14mu} \mathfrak{so}(8,2) \, \rightarrow\, \mathfrak{su}(4,1) \, \rightarrow\,
                                        \mathfrak{su}(2,1) \cr
         & \\[3pt]
         & \cr
         &\mathfrak{f}_{4(4)} \, \rightarrow\, \mathfrak{so}(4,4) \, \rightarrow\,
                                        \mathfrak{g}_{2(2)} \, \rightarrow\, \mathfrak{sl}(3,\mathbb{R})
    \end{aligned}
\end{equation}

In this paper we shall restrict ourselves to truncations to
subalgebras that have rank 3 or higher. The minimal unitary
realizations of the rank two Lie groups $\mathrm{G}_{2(2)}$ and
$\mathrm{SL}\left(3,\mathbb{R}\right)$ will be given elsewhere \cite{gpfuture}. The
minimal unitary realization of $\mathrm{SU}(2,1)$ was given in
\cite{GKN:2}.

\subsection{Truncation to the minimal unitary realization of  $\mathfrak{e}_{7(-5)}$ as a quasiconformal subalgebra}

 In order to truncate the above minimal unitary realization of $
\mathfrak{e}_{8(-24)}$ down to its subalgebra $\mathfrak{e}_{7(-5)}$
whose maximal compact subalgebra is $\mathfrak{so}^*(12) \oplus
\mathfrak{su}(2)$  we first observe that $\mathfrak{e}_{7(-5)}$ has
the 5-grading
\begin{equation}
 \mathfrak{e}_{7(-5)} \mspace{5mu} = \mspace{5mu}
\begin{array}{ccccccccc}
   \mathfrak{g}^{-2} & \oplus & \mathfrak{g}^{-1} & \oplus & \mathfrak{g}^{0} & \oplus &
   \mathfrak{g}^{+1} & \oplus & \mathfrak{g}^{+2} \\
   \mathbf{1} & \oplus & \mathbf{32} & \oplus & \left( \mathfrak{so}^\ast(12)\oplus \mathbf{1} \right) & \oplus &
   \mathbf{32} & \oplus & \mathbf{1}
\end{array}
\end{equation}
Furthermore, we note that $\mathfrak{e}_{7(-25)}$ has a subalgebra
$ \mathfrak{so}^\ast(12) \oplus \mathfrak{su}(2)$.
 Hence $\mathfrak{e}_{7(-5)}$ is centralized by an
$\mathfrak{su}(2)$ subalgebra, which can be identified  with the
one in $\mathfrak{su}(6) \oplus \mathfrak{su}(2) \oplus
\mathfrak{u}(1) \subset \mathfrak{su}(6,2) \subset
\mathfrak{e}_{7(-25)}$. Under the subalgebra $\mathfrak{su}(6)$
the adjoint $ \mathbf{66}$ and the spinor representation
$\mathbf{32}$ of $ \mathfrak{so}^\ast(12)$ decompose as follows:
\begin{equation*}
   \mathbf{32} = \mathbf{15}  \oplus \mathbf{1} \oplus \overline{\mathbf{15}} \oplus \mathbf{1} \qquad\text{and}\qquad
   \mathbf{66} = \mathbf{35} \oplus \mathbf{15} \oplus \overline{\mathbf{15}} \oplus \mathbf{1}\,.
\end{equation*}

This truncation is thus implemented by setting
\begin{equation*}
\begin{split}
   \Tilde{Z}_{7b} = 0 \text{ and } Z^{7b} = 0 \text{ where } b \not= 8 \,, \\
   \Tilde{Z}_{7b} = 0 \text{ and } Z^{8b} = 0 \text{ where } b \not= 7 \,,
\end{split}
\end{equation*}
i.e. by restricting to the $\mathfrak{su}(2)$ singlet sector.

For the sake of notational convenience, we would retain symbols
$Z^{ab}$ and $\Tilde{Z}_{ab}$ to denote creation and annihilation
operators transforming as $\mathbf{15}$ and
$\overline{\mathbf{15}}$ of $\mathfrak{su}(6) \subset
\mathfrak{so}^\ast(12)$, where $a$ and $b$ now run from $1$ to
$6$. Then, generators in $\mathfrak{g}^{-1}\oplus
\mathfrak{g}^{-2}$ of $\mathfrak{e}_{7(-5)}$ are given as follows
\begin{equation}
\begin{split}
  E = \frac{1}{2}y^2  \quad
  E^{ab} = y Z^{ab} \quad E^+ = y Z^{78} \quad
  \Tilde{E}_{ab} = y \Tilde{Z}_{ab} \quad
  E_- = y \Tilde{Z}_{78}
\end{split}
\end{equation}
The grade zero generators are $\Delta$ and
\begin{equation}
\begin{split}
   {J^{a}}_b &= 2 {Z}^{ac} \Tilde{Z}_{bc} - \frac{1}{3} {\delta^a}_b {Z}^{cd} \Tilde{Z}_{cd} \\
   J^{ab} &= \frac{1}{6} {Z}^{ab} {Z}^{78} -
                  \frac{1}{48} \epsilon^{abefgh} \Tilde{Z}_{ef} \Tilde{Z}_{gh} \\
   \Tilde{J}_{ab} &= - \frac{1}{6} \Tilde{Z}_{ab} \Tilde{Z}_{78} + \frac{1}{48} \epsilon_{abefgh} Z^{ef} Z^{gh} \\
   H &= -\frac{1}{4} \left( Z^{78}\Tilde{Z}_{78} + \Tilde{Z}_{78} Z^{78}\right) +
         \frac{1}{24}  \left(  Z^{ab} \Tilde{Z}_{ab} + \Tilde{Z}_{ab} Z^{ab} \right)
\end{split}
\end{equation}
which form the $\mathfrak{so}^\ast(12)$ subalgebra. They satisfy the following commutation relations
\begin{equation}
\begin{split}
\begin{aligned}
  \left[ {J^a}_b \,, {J^c}_d \right] &= {\delta^c}_b {J^a}_d - {\delta^d}_a {J^c}_b  \\
  \left[ {J^a}_b \,, J^{cd} \right] & = - 2 {\delta^{[c}}_b J^{d]a} - \frac{1}{3} {\delta^a}_b J^{cd} \\
  \left[ {J^a}_b \,, \Tilde{J}_{cd} \right] & =  2 {\delta^{a}}_{[c} \Tilde{J}_{d]b} +
                                            \frac{1}{3} {\delta^a}_b \Tilde{J}_{cd} \\
\end{aligned}
\begin{aligned}
  \left[ J^{ab} \,, \Tilde{J}_{cd} \right] &=
                \frac{1}{18} \left( 2 {\delta^{[a}}_{[c} {J^{b]}}_{d]} - \delta^{ab}_{cd} H \right) \\
  \left[ H \,, \Tilde{J}_{ab} \right] &= -\frac{1}{6} \Tilde{J}_{ab} \\
  \left[ H \,, J^{ab} \right] & = \frac{1}{6} J^{ab}
\end{aligned}
\end{split}
\end{equation}
In order to construct positive grade generator we need quadratic Casimir of
$\mathfrak{so}^\ast(12)$:
\begin{equation}
\begin{split}
  \mathcal{C}_2 \left( \mathfrak{so}^\ast(12) \right) &=
   \frac{1}{6} {J^a}_b {J^b}_a + 4 H^2 + 24 \left( J^{ab} \Tilde{J}_{ab} + \Tilde{J}_{ab} J^{ab} \right) =
   I_4 - \frac{99}{16} = \\
   &= \frac{1}{2} \left( \Tilde{Z}_{ab} Z^{bc} \Tilde{Z}_{cd} Z^{da} +
                     Z^{ab} \Tilde{Z}_{bc} Z^{cd} \Tilde{Z}_{da} \right)\\
      & + \frac{1}{2} \left( Z^{78} \Tilde{Z}_{78}  Z^{78} \Tilde{Z}_{78} +
           \Tilde{Z}_{78} Z^{78} \Tilde{Z}_{78}  Z^{78}  \right)+ \\
       &  - \frac{1}{8} \left( \Tilde{Z}_{ab} Z^{ab} \Tilde{Z}_{cd} Z^{cd} +
                  Z^{ab} \Tilde{Z}_{ab}  Z^{cd} \Tilde{Z}_{cd}  \right) + 4 \\
      & - \frac{1}{4} \left( Z^{ab} \Tilde{Z}_{ab} Z^{78} \Tilde{Z}_{78} +
               Z^{78} \Tilde{Z}_{78} Z^{ab} \Tilde{Z}_{ab} \right) \\
     &  - \frac{1}{4} \left(
                  \Tilde{Z}_{ab} Z^{ab} \Tilde{Z}_{78} Z^{78}  +
                \Tilde{Z}_{78} Z^{78} \Tilde{Z}_{ab} Z^{ab}  \right) \\
      &   + \frac{1}{12} \epsilon_{abcdef} Z^{ab} Z^{cd} Z^{ef} Z^{78}
        + \frac{1}{12} \epsilon^{abcdef} \Tilde{Z}_{ab}
                          \Tilde{Z}_{cd} \Tilde{Z}_{ef} \Tilde{Z}_{78}
\end{split}
\end{equation}
where the quartic invariant is built out of the spinor
representation $\mathbf{32}$ of $\mathfrak{so}^\ast(12)$. Then
generators of $\mathfrak{g}^{+1}$ are defined via
\eqref{eq:su62grade1}. Commutation relations of $\mathfrak{g}^0$
with $\mathfrak{g}^{-1}$ read
\begin{equation*}
\begin{split}
   \left[ {J^a}_b \,, E^{cd} \right] &= - 2 {\delta^{[c}}_b E^{d]a} - \frac{1}{3} {\delta^a}_b E^{cd} \qquad
   \left[ {J^a}_b \,, \Tilde{E}_{cd} \right] =  2 {\delta^{a}}_{[c} \Tilde{E}_{d]a} + \frac{1}{3} {\delta^a}_b \Tilde{E}_{cd} \\
   \left[ J^{ab} \,, E^{cd} \right] &= - \frac{1}{24} \epsilon^{abcdef} \Tilde{E}_{ef} \qquad
     \left[ \Tilde{J}_{ab} \,, E^{cd} \right] =  - \frac{1}{6} \delta^{cd}_{ab} \Tilde{E}_{78} \\
    \left[ \Tilde{J}_{ab} \,, \Tilde{E}_{cd} \right] &= - \frac{1}{24} \epsilon_{abcdef} E^{ef} \qquad
    \left[ J^{ab} \,, \Tilde{E}_{cd} \right] = - \frac{1}{6} \delta^{ab}_{cd} E^{78}
\end{split}
\end{equation*}
\begin{equation*}
\begin{split}
    \left[ H \,, E^{ab} \right] &= \frac{1}{12} E^{ab} \qquad
       \left[ H \,, \Tilde{E}_{ab} \right] = - \frac{1}{12} \Tilde{E}_{ab} \\
    \left[ J^{ab} \,, E^{78} \right] &= 0 \qquad
     \left[ \Tilde{J}_{ab} \,, E^{78} \right] = - \frac{1}{12} \Tilde{E}_{ab} \qquad
     \left[ H \,, E^{78} \right] = - \frac{1}{4} E^{78}  \\
     \left[ \Tilde{J}_{ab} \,, \Tilde{E}_{78} \right] & = 0 \qquad
     \left[ J^{ab} \,, \Tilde{E}_{78} \right] = - \frac{1}{12} E^{ab} \qquad
     \left[ H \,, \Tilde{E}_{78} \right] = + \frac{1}{4} \Tilde{E}_{78}
\end{split}
\end{equation*}
Commutators of $\mathfrak{so}^\ast(12)$ generators and the
generators belonging to $\mathfrak{g}^{+1}$ subspace are obtained
by substituting $E^{ab}$ with $F^{ab}$ and $\tilde{E}_{ab}$ with
$\tilde{F}_{ab}$ in equations above. Spaces $\mathfrak{g}^{\pm 2}$
are $\mathfrak{so}^\ast(12)$ singlets each. Elements of
$\mathfrak{g}^{\pm 2}$ together with $\Delta$ generate an
$\mathfrak{sl}(2, \mathbb{R}) \subset \mathfrak{e}_{7(-5)}$
subalgebra
\begin{equation}
   \left[ E, F \right] = - \Delta \qquad
   \left[ \Delta \,, E \right] = - 2 E \qquad
   \left[ \Delta \,, F \right] = + 2 F \,.
\end{equation}
Generators in $\mathfrak{g}^{-1}$ and $\mathfrak{g}^{+1}$ close into $\mathfrak{g}^{0}$ as follows
\begin{equation}
\begin{split}
 \left[ E^{ab} \,, F^{cd} \right] &= -6 i \epsilon^{abcdef} \Tilde{J}_{ef} \\
 \left[ E^{ab} \,, \Tilde{F}_{cd} \right] &= - i \delta^{ab}_{cd} \left( 4 H - \Delta \right) - 4 i {\delta^{[a}}_{[c} {J^{b]}}_{d]} \\
 \left[ E^{ab} \,, F^{78} \right] &= -12 i J^{ab}   \qquad
 \left[ E^{ab} \,, \Tilde{F}_{78} \right] = 0 \\
\end{split}
\end{equation}
\begin{equation}
\begin{split}
  \left[ \Tilde{E}_{ab} \,, F^{cd} \right] &= - i \delta^{cd}_{ab} \left( 4 H + \Delta \right) - 4 i {\delta^{[c}}_{[a} {J^{d]}}_{b]} \\
  \left[ \Tilde{E}_{ab} \,, \Tilde{F}_{cd} \right] &= + 6 i \epsilon_{abcdef} J^{ef} \\
  \left[ \Tilde{E}_{ab} \,, \Tilde{F}_{78} \right] &= +12 i \Tilde{J}_{ab} \qquad
    \left[ \Tilde{E}_{ab} \,, F^{78} \right] = 0
\end{split}
\end{equation}
\begin{equation}
\begin{split}
  \left[ E^{78} \,, F^{ab} \right] &= - 12 i J^{ab} \quad
  \left[ E^{78} \,, \Tilde{F}_{78} \right] = i \left( \frac{1}{2} \Delta + 6 H \right)  \\
    \left[ E^{78} \,, \Tilde{F}_{ab} \right] &= 0 \qquad
  \left[ E^{78} \,, F^{78} \right] = 0 \\
  \left[ \Tilde{E}_{78}\,, F^{ab} \right] &= 0 \qquad
    \left[ \Tilde{E}_{78}\,, \Tilde{F}_{78} \right] = 0 \\
  \left[ \Tilde{E}_{78}\,, F^{78} \right] &= i \left( - \frac{1}{2} \Delta + 6 H \right) \quad
  \left[ \Tilde{E}_{78}\,, \Tilde{F}_{ab} \right] = + 12 i \Tilde{J}_{ab}
\end{split}
\end{equation}
The resulting realization of $\mathfrak{e}_{7(-5)}$ is that of the
minimal unitary representation  and the quadratic Casimir of
$\mathfrak{e}_{7(-5)}$  reduces to a c-number as required by
irreducibility of the minimal unitary representation
\begin{equation}
\begin{split}
  \mathcal{C}_2 \left( \mathfrak{e}_{7(-5)} \right) &= \mathcal{C}_2 \left( \mathfrak{so}^\ast(12) \right) +
               \frac{1}{12} \Delta^2 + \frac{1}{6} \left( F E + E F \right) \\
                & - \frac{1}{12} \left( \Tilde{E}_{ab} F^{ab} +  F^{ab} \Tilde{E}_{ab}
                                          - \Tilde{F}_{ab} E^{ab} - E^{ab} \Tilde{F}_{ab} \right)   \\
               & - \frac{1}{6} \left( \Tilde{E}_{78} F^{78} +  F^{78} \Tilde{E}_{78}
                                          - \Tilde{F}_{78} E^{78} - E^{78} \Tilde{F}_{78} \right) \\
               & = \left( I_4 -\frac{99}{16} \right) + \left( \frac{1}{3} I_4 - \frac{1}{16} \right) +
                   \left( - \frac{4}{3} I_4 - \frac{31}{4} \right) = - 14
\end{split}
\end{equation}

\subsection{Truncation to the minimal unitary realization of $\mathfrak{e}_{6(2)}$ as a quasiconformal subalgebra}

Quasi-conformal algebra $\mathfrak{e}_{6(2)}$ with the maximal
compact subalgebra $ \mathfrak{su(6)} \oplus \mathfrak{su(2)} $
has the following 5-graded decomposition
\begin{equation*}
   \mathbf{78} = \mathbf{1} \oplus \mathbf{20}
                \oplus \left( \mathfrak{su}(3,3) \oplus \Delta \right)
                \oplus \mathbf{20} \oplus \mathbf{1}
\end{equation*}
and since $\mathfrak{su}(3,3) \subset \mathfrak{so}^\ast(12)$ it
can be obtained by the further truncation of
$\mathfrak{e}_{7(-5)}$. The maximal compact subalgebra
$\mathfrak{su}(3) \oplus \mathfrak{su}(3) \oplus \mathfrak{u}(1) $
of $\mathfrak{su}(3,3)$  is also a subalgebra of $
\mathfrak{su}(6) \subset \mathfrak{so}^\ast(12) $. This suggests
that we split $\mathfrak{su}(6)$ indices $a =1, \dots , 6$ into
two subsets, $\Check{a} =(1,2,3)$ and $\Hat{a}= (4,5,6)$, and keep
only oscillators which have both types of indices in addition to
singlets $\Tilde{Z}_{78}$ and $Z^{78}$, i.e. set
\begin{equation}
  Z^{\Check{a}\Check{c}} = 0 \qquad \Tilde{Z}_{\Check{a}\Check{c}} = 0 \qquad
  Z^{\Hat{a}\Hat{c}} = 0 \qquad \Tilde{Z}_{\Hat{a}\Hat{c}} = 0
\end{equation}
Indeed corresponding $\mathfrak{su}(3) \oplus \mathfrak{su}(3) \subset \mathfrak{su}\left(3,3\right)$ branching
reads
\begin{equation*}
   \mathbf{20} = \left( \mathbf{1},\, \mathbf{1} \right) \oplus
                 \left( \mathbf{3},\, \mathbf{3} \right) \oplus
         \left( \overline{\mathbf{3}} ,\, \overline{\mathbf{3}} \right) \oplus
         \left( \mathbf{1},\, \mathbf{1} \right)
\end{equation*}
This reduction is quite straightforward, and we shall not give
here complete commutation relations.  All of the formulae of
$\mathfrak{e}_{7(-5)}$ carry over to this case provided we set to
zero appropriate operators. The quadratic Casimir of
$\mathfrak{su}(3,3)$ that is needed  to construct generators of
$\mathfrak{g}^{+1} \oplus \mathfrak{g}^{+2}$ reads as follows
\begin{equation}
\begin{split}
 & \mathcal{C}_2 \left( \mathfrak{su}\left(3,\,3\right) \right) = \frac{1}{6} {J^{\Check{a}}}_{\Check{c}} {J^{\Check{c}}}_{\Check{a}} +
    \frac{1}{6} {J^{\Hat{a}}}_{\Hat{c}} {J^{\Hat{c}}}_{\Hat{a}} + 4 H^2
            + 24 \left( J^{\Check{a}\Check{c}} \Tilde{J}_{\Check{a}\Check{c}} +
               \Tilde{J}_{\Check{a}\Check{c}} J^{\Check{a}\Check{c}}\right) \\
        & +  24 \left( J^{\Hat{a}\Hat{c}} \Tilde{J}_{\Hat{a}\Hat{c}} +
                \Tilde{J}_{\Hat{a}\Hat{c}} J^{\Hat{a}\Hat{c}}\right) = I_4 - \frac{35}{16} =
         - \frac{1}{8} \left( \Tilde{Z}_{ab} Z^{ab} \Tilde{Z}_{cd} Z^{cd} +
                  Z^{ab} \Tilde{Z}_{ab}  Z^{cd} \Tilde{Z}_{cd}  \right)   \\
   &+ \frac{1}{2} \left( \Tilde{Z}_{ab} Z^{bc} \Tilde{Z}_{cd} Z^{da} +
                     Z^{ab} \Tilde{Z}_{bc} Z^{cd} \Tilde{Z}_{da} \right)
       + \frac{1}{2} \left( Z^{78} \Tilde{Z}_{78}  Z^{78} \Tilde{Z}_{78} +
           \Tilde{Z}_{78} Z^{78} \Tilde{Z}_{78}  Z^{78}  \right) \\
      & - \frac{1}{4} \left( Z^{ab} \Tilde{Z}_{ab} Z^{78} \Tilde{Z}_{78} +
               Z^{78} \Tilde{Z}_{78} Z^{ab} \Tilde{Z}_{ab} \right)
       - \frac{1}{4} \left(
                  \Tilde{Z}_{ab} Z^{ab} \Tilde{Z}_{78} Z^{78}  +
                \Tilde{Z}_{78} Z^{78} \Tilde{Z}_{ab} Z^{ab}  \right) \\
      &   + \frac{1}{12} \epsilon_{abcdef} Z^{ab} Z^{cd} Z^{ef} Z^{78}
        + \frac{1}{12} \epsilon^{abcdef} \Tilde{Z}_{ab}
                          \Tilde{Z}_{cd} \Tilde{Z}_{ef} \Tilde{Z}_{78} + \frac{5}{4}
\end{split}
\end{equation}
where $Z^{ab}$ and $\Tilde{Z}_{ab}$ are as described above, and
hence $I_4$ is the quadratic invariant of $\mathfrak{su}(3,3)$ in
the representation $\mathbf{20}$. The resulting realization of
$\mathfrak{e_{6(2)}}$ is again that of the minimal unitary
representation.   Because some of the oscillators were set equal
to zero in the truncation, they do not contribute to the value of
the quadratic Casimir of the algebra, the c-number to which it
reduces is now different
\begin{equation}
   \mathcal{C}_2 \left( \mathfrak{e}_{6(2)} \right) = \left( I_4 - \frac{35}{16} \right) +
      \left( \frac{1}{3} I_4 - \frac{1}{16} \right) + \left( - \frac{4}{3} I_4 - \frac{15}{4} \right) = - 6
\end{equation}

\subsection{Truncation to the minimal unitary realization of  $\mathfrak{e}_{6(-14)}$ as a quasiconformal subalgebra}
Quasiconformal realization of another real form  of
$\mathfrak{e}_6$ , namely $\mathfrak{e}_{6(-14)}$ with the maximal
compact subalgebra $ \mathfrak{so(10)} \oplus \mathfrak{so(2)} $ ,
can also be obtained by truncation of $\mathfrak{e}_{7(-5)}$. Its
five-graded decomposition reads as follows
\begin{equation*}
   \mathfrak{e}_{6(-14)} = \mathbf{1} \oplus \mathbf{20}
                \oplus \left( \mathfrak{su}(5,1) \oplus \Delta \right)
                \oplus \mathbf{20} \oplus \mathbf{1}
\end{equation*}
In order to implement this truncation we observe the following chain of inclusions
\begin{equation*}
   \mathfrak{su}(5,1) \subset \mathfrak{so}(10,2) \subset \mathfrak{e}_{7(-25)}
\end{equation*}
Subalgebra $\mathfrak{so}(10,2)$ is centralized by $\mathfrak{su}(1,1)$
while $\mathfrak{su}(5,1)$ is centralized by $\mathfrak{su}(2,1)$ within $\mathfrak{e}_{7(-25)}$,
suggesting that we only keep oscillators $Z^{ab}$ and $\Tilde{Z}_{ab}$
with indexes now running from 1 to 5 as follows from $\mathfrak{u}(5) \subset \mathfrak{su}(5,1)$
branching of $\mathbf{20} = \mathbf{10} \oplus \overline{\mathbf{10}}$.
Then generators of $\mathfrak{su}(5,1)$ are given as follows
\begin{equation}
\begin{split}
  {J^a}_b &= 2 Z^{ac} \Tilde{Z}_{bc} - \frac{2}{5} {\delta^a}_b Z^{cd} \Tilde{Z}_{cd} \qquad
  H = \frac{1}{24} \left( Z^{ab} \Tilde{Z}_{ab} + \Tilde{Z}_{ab} Z^{ab} \right) \\
  J^a &= -\frac{1}{48} \epsilon^{abcde} \Tilde{Z}_{bc} \Tilde{Z}_{de} \qquad
  \Tilde{J}_a = + \frac{1}{48} \epsilon_{abcde} {Z}^{bc} {Z}^{de}
\end{split}
\end{equation}
with commutation relations
\begin{equation}
\begin{split}
  \left[ {J^a}_b \,, {J^c}_d \right] &= {\delta^c}_b {J^a}_d - {\delta^d}_a {J^c}_b \qquad
  \left[ {J^a}_b \,, J^c \right] = {\delta^c}_b J^a - \frac{1}{5} {\delta^a}_b J^c \\
  \left[ {J^a}_b \,, \Tilde{J}_c \right] &= - {\delta^a}_c \Tilde{J}_b + \frac{1}{5} {\delta^a}_b \Tilde{J}_c \quad
  \left[ H, J^a \right] = - \frac{1}{6} J^a \quad
   \left[ H, \Tilde{J}_a \right] = + \frac{1}{6} \Tilde{J}_a \\
  \left[ J^a \,, \Tilde{J}_b \right] &= \frac{1}{144} {J^a}_b - \frac{1}{20} {\delta^a}_b H \qquad
   \left[ J^a \,, J^b \right] = 0 \qquad \left[ \Tilde{J}_a \,, \Tilde{J}_b \right] = 0
\end{split}
\end{equation}
resulting in the following Casimir
\begin{equation}
\begin{split}
  \mathcal{C}_2 \left( \mathfrak{su}(5,1) \right) &=
        \frac{1}{6} {J^a}_b  {J^b}_a + \frac{36}{5} H^2 +
           24 \left( J^a \Tilde{J}_a + \Tilde{J}_a J^a \right) = I_4 - \frac{35}{16} \\
    &= \frac{1}{2} \left( \Tilde{Z}_{ab} Z^{bc} \Tilde{Z}_{cd} Z^{da} +
                     Z^{ab} \Tilde{Z}_{bc} Z^{cd} \Tilde{Z}_{da} \right)\\
    &- \frac{1}{8} \left( \Tilde{Z}_{ab} Z^{ab} \Tilde{Z}_{cd} Z^{cd} +
                  Z^{ab} \Tilde{Z}_{ab}  Z^{cd} \Tilde{Z}_{cd}  \right) + \frac{5}{4}
\end{split}
\end{equation}
Remaining generators of $\mathfrak{e}_{6(-14)}$ and their
commutators straight\-forwardly follow from those of
$\mathfrak{e}_{7(-5)}$. We shall only present the c-number to
which the quadratic Casimir of $\mathfrak{e}_{6(-14)}$ reduces
upon evaluation on the resulting minimal unitary realization
\begin{equation}
  \mathcal{C}_2 \left( \mathfrak{e}_{6(-14)} \right) = \left( I_4 - \frac{35}{16} \right) +
     \left( \frac{1}{3} I_4 - \frac{1}{16} \right) + \left( - \frac{4}{3} I_4 - \frac{15}{4} \right) = - 6
\end{equation}

\subsection{Truncation to the minimal unitary realization of  $\mathfrak{f}_{4(4)}$ as a quasiconformal subalgebra}

The realization of the Lie algebra $\mathfrak{e}_{6(2)}$ given above can be further truncated to obtain the
minimal unitary realization of the Lie algebra $\mathfrak{f}_{4(4)}$ with the maximal compact subalgebra $
\mathfrak{usp(6)} \oplus \mathfrak{usp(2)}$.
 The five graded structure of $\mathfrak{f}_{4(4)}$ as a quasiconformal algebra reads as follows
\begin{equation}
   \mathbf{52}=\mathfrak{f}_{4(4)} = \mathbf{1} \oplus \mathbf{14} \oplus
           \left( \mathfrak{sp}\left( 6 \,, \mathbb{R} \right) \oplus \Delta \right) \oplus
              \mathbf{14} \oplus \mathbf{1}
\end{equation}
One way to obtain the truncation of $\mathfrak{e}_{6(2)}$  to $\mathfrak{f}_{4(4)}$ is suggested by
$\mathfrak{u}(3) \subset \mathfrak{sp}\left( 6,\, \mathbb{R}\right)$ branching of $\mathbf{14} = \mathbf{1}
\oplus \mathbf{6} \oplus \overline{\mathbf{6}} \oplus \mathbf{1}$. It amounts to identifying
\void{$Z^{\Check{a}\Hat{c}} \sim Z^{\Check{c}\Hat{a}}$ } the two $\mathfrak{su}(3)$ sub\-algebra of
$\mathfrak{su}(3,3) \subset \mathfrak{e}_{6(2)}$ and discarding the antisymmetric components $Z^{[ab]}$ of
$Z^{ab}$ .

Let us define the  symmetric tensor oscillators $S^{ac} = Z^{(ac)}$ and $\Tilde{S}_{ac} = \Tilde{Z}_{(ac)}$, $
a,b,...=1,2,3$,  which correspond to independent oscillators left after the identification. They satisfy the
following commutation relations
\begin{equation}
   \left[ \Tilde{S}_{cd} \,, S^{ab} \right] = \frac{1}{4} \left( {\delta^a}_c {\delta^b}_d +
         {\delta^b}_c {\delta^a}_d \right)
\end{equation}
With these oscillators we build generators of $\mathfrak{sp}\left(6\,, \mathbb{R}\right)$:
\begin{equation}
\begin{split}
   {J^a}_b &= 2 S^{ac} \Tilde{S}_{bc} - \frac{2}{3} {\delta^a}_b S^{cd} \Tilde{S}_{cd} \\
   H &= - \frac{1}{4} \left( Z^{78} \Tilde{Z}_{78} +  \Tilde{Z}_{78} Z^{78} \right) +
         \frac{1}{12} \left( S^{ab} \Tilde{S}_{ab} +\Tilde{S}_{ab} S^{ab} \right) \\
  J^{ab} &= \frac{1}{6} S^{ab} Z^{78} + \frac{1}{12}  \epsilon^{acd} \epsilon^{bef} \Tilde{S}_{ce} \Tilde{S}_{df} \\
  \Tilde{J}_{ab} &= -\frac{1}{6} \Tilde{S}_{ab} \Tilde{Z}_{78} -
                \frac{1}{12} \epsilon_{acd} \epsilon_{bef} S^{ce} S^{df}
\end{split}
\end{equation}
satisfying the following commutation relations
\begin{equation}
\begin{aligned}
  &\left[ {J^a}_b \,, J^{cd} \right] =  {\delta^{(c}}_b J^{d)a} - \frac{1}{3} {\delta^a}_b J^{cd} \\
  &\left[ {J^a}_b \,, \Tilde{J}_{cd} \right]  =  - {\delta^{a}}_{(c} \Tilde{J}_{d)b}
                                  + \frac{1}{3} {\delta^a}_b \Tilde{J}_{cd} \\
  &\left[ J^{ab} \,, \Tilde{J}_{cd} \right] = \frac{1}{72} \left(
             {\delta^{(a}}_{(c} {J^{b)}}_{d)} -
             2 {\delta^{(a}}_{(c} {\delta^{b)}}_{d)} H \right)
\end{aligned}
 \quad
\begin{aligned}
  &\left[ {J^a}_b \,, {J^c}_d \right] = {\delta^c}_b {J^a}_d - {\delta^d}_a {J^c}_b \\
  &\left[ H \,, J^{ab} \right] = - \frac{1}{6} J^{ab} \\
  &\left[ H \,, \Tilde{J}_{ab} \right] =  \frac{1}{6} \Tilde{J}_{ab}
\end{aligned}
\end{equation}
The quadratic Casimir of $\mathfrak{sp}(6,\mathbb{R})$ is then given by
\begin{equation}
\begin{split}
  \mathcal{C}_2\left(\mathfrak{sp}(6,\mathbb{R}) \right) &=
       \frac{1}{3}\, {J^a}_b  {J^b}_a + 4 \, H^2 + 24 \left( J^{ab} \Tilde{J}_{ab} + \Tilde{J}_{ab} J^{ab} \right) =
           I_4 - \frac{15}{16} \\
          &=  \left( \Tilde{S}_{ab} S^{bc} \Tilde{S}_{cd} S^{da} +
                     S^{ab} \Tilde{S}_{bc} S^{cd} \Tilde{S}_{da} \right)\\
      & + \frac{1}{2} \left( Z^{78} \Tilde{Z}_{78}  Z^{78} \Tilde{Z}_{78} +
           \Tilde{Z}_{78} Z^{78} \Tilde{Z}_{78}  Z^{78}  \right)+ \\
       &  - \frac{1}{2} \left( \Tilde{S}_{ab} S^{ab} \Tilde{S}_{cd} S^{cd} +
                  S^{ab} \Tilde{S}_{ab}  S^{cd} \Tilde{S}_{cd}  \right) + \frac{7}{16} \\
      & - \frac{1}{2} \left( S^{ab} \Tilde{S}_{ab} Z^{78} \Tilde{Z}_{78} +
               Z^{78} \Tilde{Z}_{78} S^{ab} \Tilde{S}_{ab} \right) \\
     &  - \frac{1}{2} \left(
                  \Tilde{S}_{ab} S^{ab} \Tilde{Z}_{78} Z^{78}  +
                \Tilde{Z}_{78} Z^{78} \Tilde{S}_{ab} S^{ab}  \right) \\
      &   - \frac{2}{3} \epsilon_{abc}\epsilon_{def} S^{ad} S^{be} S^{cf} Z^{78}
        - \frac{2}{3} \epsilon^{abc}\epsilon^{def} \Tilde{S}_{ad}
                          \Tilde{S}_{be} \Tilde{S}_{cf} \Tilde{Z}_{78}
\end{split}
\end{equation}
Negative grade generators are defined as
\begin{equation}
   E = \frac{1}{2} y \quad
   E^{ab} = y S^{ab} \quad E^{+} = y Z^{78} \quad
   \Tilde{E}_{ab} = y \Tilde{S}_{ab} \quad E_{-} = y \Tilde{Z}_{78}
\end{equation}
They satisfy commutation relations, different from those of negative grade
generators of $\mathfrak{e}_{7(-5)}$
\begin{equation}
   \left[ \Tilde{E}_{ab} \,, E^{cd} \right] = {\delta^{(c}}_{(b} {\delta^{b)}}_{d)} E
\end{equation}
reflecting that $S^{ab}$ and $\Tilde{S}_{ab}$ are now symmetric tensor oscillators. Positive grade generators,
and their commutator, are given by the following equations
\begin{equation}
\begin{aligned}
   F &= \frac{1}{2} p^2 + 2 i y^{-2} I_4 \\
   F^{ab} &= - p S^{ab} + 2 i y^{-1} \left[ S^{ab} \,, I_4 \right] \\
   \Tilde{F}_{ab} &= - p \Tilde{S}_{ab} + 2 i y^{-1} \left[  \Tilde{S}_{ab} \,, I_4 \right]
\end{aligned} \quad\qquad
  \begin{aligned}
   \left[ \Tilde{F}_{ab} \,, F^{cd} \right] = {\delta^{(c}}_{(b} {\delta^{b)}}_{d)} F
  \end{aligned}
\end{equation}
Quadratic Casimir of the resulting minimal unitary realization of $\mathfrak{f}_{4(4)}$
\begin{equation}
\begin{split}
  \mathcal{C}_2 \left( \mathfrak{f}_{4(4)} \right) &= \mathcal{C}_2 \left( \mathfrak{sp}(6, \mathbb{R}) \right)
             + \frac{1}{12} \Delta^2 + \frac{1}{6} \left( F E + E F \right) \\
        & + i \left( \Tilde{E}_{ab} F^{ab} + F^{ab} \Tilde{E}_{ab}
                     - \Tilde{F}_{ab} E^{ab} - E^{ab} \Tilde{F}_{ab} \right) \\
       & - \frac{i}{6} \left( \Tilde{E}_{78} F^{78} + F^{78} \Tilde{E}_{78}
                 - \Tilde{F}_{78} E^{78} - E^{78} \Tilde{F}_{78} \right)
\end{split}
\end{equation}
reduces to a c-number
\begin{equation}
 \mathcal{C}_2 \left( \mathfrak{f}_{4(4)} \right) = \left( I_4 - \frac{15}{16} \right) +
     \left( \frac{1}{3} I_4 - \frac{1}{16} \right) + \left( - \frac{4}{3} I_4 - \frac{9}{4} \right)  = - \frac{13}{4}
\end{equation}
in agreement with parent algebras and as required by  irreducibility.

\subsection{Truncation to the minimal unitary realization \\ 
   of  $\mathfrak{so}\left(4,4\right)$ as a quasiconformal subalgebra}
We further truncate $\mathfrak{f}_{4(4)}$ to obtain the minimal unitary realization of
$\mathfrak{so}\left(4,4\right)$ which has the following 5-graded decomposition
\begin{equation}
   \mathbf{28} = \mathbf{1} \oplus \left( \mathbf{2},  \mathbf{2},  \mathbf{2} \right) \oplus \left(
   \mathfrak{sp}\left(2,\,\mathbb{R} \right) \oplus
   \mathfrak{sp}\left(2,\,\mathbb{R} \right) \oplus
   \mathfrak{sp}\left(2,\,\mathbb{R} \right) \oplus
   \Delta \right) \oplus  \left( \mathbf{2},  \mathbf{2},  \mathbf{2} \right) \oplus \mathbf{1}
\end{equation}
This truncation is achieved by restricting  $S^{ab}$ and $\Tilde{S}_{ab}$ operators to their  diagonal components
\begin{equation}
   S^{ab} = \delta^{ab} S^a \qquad \Tilde{S}_{ab} = \delta_{ab} \Tilde{S}_a \qquad
   \left[ \Tilde{S}_a ,\, S^b \right] = \frac{1}{2} {\delta^b}_a
\end{equation}
where $a,b,..=1,2,3$, and discarding the off-diagonal oscillators.  Three copies of
$\mathfrak{sp}\left(2,\,\mathbb{R}\right)$ are generated by
\begin{equation}
  \begin{aligned}
     J^a_- &= -\frac{1}{6} \Tilde{S}_a \Tilde{Z}_{78} - \frac{1}{12} \epsilon_{abc} S^b S^c \cr
     J^a_+ &= \frac{1}{6} S^a Z^{78} + \frac{1}{12} \epsilon^{abc} \Tilde{S}_b \Tilde{S}_c \cr
  \end{aligned}
\quad
     J^a_0 = Z^{78} \Tilde{Z}_{78} + \sum_{b=1}^3 \left( 2 \delta^{ab} - 1 \right) S^b \Tilde{S}_b
\end{equation}
The quadratic Casimir of $\mathfrak{sp}\left(2,\,\mathbb{R} \right) \oplus
   \mathfrak{sp}\left(2,\,\mathbb{R}\right) \oplus
   \mathfrak{sp}\left(2,\,\mathbb{R} \right)$ then reads
\begin{equation}
\begin{aligned}
   \mathcal{C}_2 &\left(    \mathfrak{sp}\left(2,\,\mathbb{R} \right) \oplus
   \mathfrak{sp}\left(2,\,\mathbb{R} \right) \oplus
   \mathfrak{sp}\left(2,\,\mathbb{R} \right) \right) = \sum_{a=1}^3 \left[ \frac{1}{3} {J^a_0}J^a_0 +
        24 \left(J^a_- J^a_+ + J^a_+ J^a_- \right) \right]\cr
   & = \sum_{a=1}^3 \left( \left( \Tilde{S}_a S^a \right)^2 +  \left(S^a \Tilde{S}_a\right)^2 \right)
+  \left( Z^{78} \Tilde{Z}_{78} \right)^2 +  \left(  \Tilde{Z}_{78} Z^{78} \right)^2 \cr
   & -
     \frac{1}{2} \left(  \sum_{a=1}^3 S^a \Tilde{S}_a + Z^{78} \Tilde{Z}_{78}\right)^2 - \frac{1}{2}
      \left(  \sum_{a=1}^3  \Tilde{S}_a S^a + \Tilde{Z}_{78} Z^{78}\right)^2 \cr
   & - 4 \, S^1 \, S^2\, S^3 \, Z^{78} - 4 \, \Tilde{S}_1\, \Tilde{S}_2\, \Tilde{S}_3 \,\Tilde{Z}_{78} = I_4 -
    \frac{3}{16}
\end{aligned}
\end{equation}
The  commutation relations of the generators in  $\mathfrak{g}^0$ are
\begin{equation}
    \left[ J^a_0 ,\, J^b_\pm \right] = \pm \delta^{ab} J^a_\pm \qquad
    \left[ J^a_+ ,\, J^b_- \right] = \frac{1}{72} \,\delta^{ab}  J^a_0
\end{equation}
Negative grade generators are
\begin{equation}
   E = \frac{1}{2} y^2 \qquad E^a = y S^a \qquad \Tilde{E}_a = y \Tilde{S}_a \qquad  E^{78}=yZ^{78} \qquad
   \Tilde{E}_{78} =y \Tilde{Z}_{78}
\end{equation}
and positive grade generators are
\begin{equation}
   F = \frac{1}{2} p^2 + 2 i y^{-2} I_4 \qquad
\begin{aligned}
  F^a &= -p S^a + 2 i y^{-1} \left[ S^a ,\, I_4 \right] \cr
  F^{78} &= -p Z^{78} + 2i y^{-1} \left[ Z^{78}, \, I_4 \right] \cr
 \Tilde{F}_a &= -p \Tilde{S}_a + 2 i y^{-1} \left[ \Tilde{S}_a ,\, I_4 \right] \cr
 \Tilde{F}_{78} &= -p \Tilde{Z}_{78} + 2i y^{-1} \left[ \Tilde{Z}_{78}, \, I_4 \right]
\end{aligned}
\end{equation}

The quadratic Casimir of $\mathfrak{so}\left(4,4\right)$
\begin{equation}
\begin{aligned}
  \mathcal{C}_2 \left( \mathfrak{so}\left(4, 4 \right) \right) &= \mathcal{C}_2 \left( \mathfrak{g}^0 \right)
  + \frac{1}{12} \Delta^2 + \frac{1}{6} \left( F E + E F \right) \cr
 &+ \frac{i}{6} \left( E^a \Tilde{F}_a + \Tilde{F}_a E^a - \Tilde{E}_a F^a - F^a \Tilde{E}_a \right) \cr &+
  \frac{i}{6} \left( E^{78} \Tilde{F}_{78} + \Tilde{F}_{78} E^{78} - F^{78} \Tilde{E}_{78} - \Tilde{E}_{78} F^{78} \right)
\end{aligned}
\end{equation}
reduces to c-number as before
\begin{equation}
   \mathcal{C}_2 \left( \mathfrak{so}\left(4, 4 \right) \right) = \left( I_4 - \frac{3}{16} \right) +
              \left( \frac{1}{3} \, I_4 - \frac{1}{16} \right) + \left( - \frac{4}{3} I_4 - \frac{13}{12}
      \right)= -\frac{4}{3}
\end{equation}

\section{Truncation to the minimal unitary realization of  $\mathfrak{e}_{7(-25)}$ as a quasiconformal subalgebra}

The group $\mathrm{E}_{7(-25)}$ has the maximal compact subgroup
$E_6 \times U(1)$ and arises as the  $U$-duality group of
exceptional $\mathcal{N}=2$ Maxwell-Einstein supergravity in $d=4$
whose scalar manifold is $\mathrm{E}_{7(-25)}/ \left( E_6 \times
U(1) \right)$. Its action on the 27 complex scalar fields can be
represented as a generalized conformal group \cite{gst,mg92,GKN:1}.
As a generalized conformal group its Lie algebra has a natural
3-graded structure
\begin{equation*}
\mathfrak{e}_{7(-25)} = \overline{27} \oplus (\mathfrak{e}_{6(-26)}
\oplus \mathfrak{so}(1,1) ) \oplus 27
\end{equation*}
The quasiconformal realization of $E_{8(-24)}$ can be truncated to
the conformal realization of $E_{7(-25)}$ in essentially two
different ways.

 In this section we will however consider a different
truncation of $E_{8(-24)}$ such that the resulting realization of
$E_{7(-25)}$ is quasiconformal corresponding to its minimal unitary
representation.

Just as the subalgebra $\mathfrak{e}_{7(-5)}$ is  normalized by
$\mathfrak{su}(2) \subset \mathfrak{su}(6,2) \subset \mathfrak{g}^0=
\mathfrak{e}_{7(-25)} $, the subalgebra $\mathfrak{e}_{7(-25)}$ is
normalized by $\mathfrak{su}(1,1) \subset \mathfrak{su}(6,2) \subset
\mathfrak{g}^0\mathfrak{e}_{7(-25)}$ within $\mathfrak{e}_{8(-24)}$.
Similarly to $\mathfrak{e}_{7(-5)}$ we obtain
\begin{equation}
   \mathfrak{e}_{7(-25)}= \mathbf{133} = \mathbf{1} \oplus \mathbf{32} \oplus \left( \mathfrak{so}(10,2) \oplus \Delta \right) \oplus \mathbf{32}
    \oplus \mathbf{1}
\end{equation}

We identify the $\mathfrak{su}(1,1)$ in question with the one generated by ${J^6}_7$, ${J^7}_6$ and ${J^6}_6-{J^7}_7$
generators of $\mathfrak{su}(6,2) \subset \mathfrak{e}_{7(-25)} \subset \mathfrak{e}_{8(-24)}$.
The truncation will then amount to setting $Z^{a6} = Z^{6a} = 0$ where $a \not= 7$, as well as $Z^{a7} = Z^{7a} = 0$ for $a \not= 6$. Let us relable coefficients and introduce $\Dot{a} = 1,\dots,5,8$. Then $\mathfrak{su}(5,1)$ is generated by
\begin{equation}
    {J^{\Dot{a}}}_{\Dot{b}} = 2 Z^{\Dot{a}\Dot{c}} \Tilde{Z}_{\Dot{b}\Dot{c}} - \frac{1}{3} {\delta^{\Dot{a}}}_{\Dot{b}}
    Z^{\Dot{d}\Dot{c}} \Tilde{Z}_{\Dot{d}\Dot{c}}
\end{equation}
The other generators of $\mathfrak{so}(10,2)$ are then given as follows
\begin{equation}
\begin{aligned}
   U &= \frac{3}{2} \left( Z^{67} \Tilde{Z}_{67} + \Tilde{Z}_{67} Z^{67} \right) - \frac{1}{4} \left( Z^{\Dot{a}\Dot{b}}
   \Tilde{Z}_{\Dot{a}\Dot{b}} + \Tilde{Z}_{\Dot{a}\Dot{b}} Z^{\Dot{a}\Dot{b}} \right) \cr
  J^-_{\Dot{a}\Dot{b}} &= - \frac{1}{6} \Tilde{Z}_{\Dot{a}\Dot{b}} \Tilde{Z}_{67} + \frac{1}{48}
   \epsilon_{\Dot{a}\Dot{b}\Dot{c}\Dot{d}\Dot{e}\Dot{f}} Z^{\Dot{c}\Dot{d}} Z^{\Dot{e}\Dot{f}} \cr
  J_+^{\Dot{a}\Dot{b}} &= \frac{1}{6} Z^{\Dot{a}\Dot{b}} Z^{67} - \frac{1}{48}
  \epsilon^{\Dot{a}\Dot{b}\Dot{c}\Dot{d}\Dot{e}\Dot{f}} \Tilde{Z}_{\Dot{c}\Dot{d}} \Tilde{Z}_{\Dot{e}\Dot{f}}
\end{aligned}
\end{equation}
satisfying the following hermiticity condition
\begin{equation}
   \left( {J^{\Dot{a}}}_{\Dot{b}} \right)^\dagger = \eta^{\Dot{a}\Dot{c}} \eta_{\Dot{b}\Dot{d}} {J^{\Dot{d}}}_{\Dot{c}}
  \qquad U^\dagger = U \qquad \left( J^-_{\Dot{a}\Dot{b}} \right)^\dagger = J_+^{\Dot{c}\Dot{d}} \eta_{\Dot{a}\Dot{c}}
   \eta_{\Dot{b}\Dot{d}}
\end{equation}
where $\eta_{\Dot{a}\Dot{b}} = \mathop\mathrm{diag}\left(+1,+1,+1,+1,+1,-1\right)$;
and the commutation relations read as follows
\begin{equation}
\begin{aligned}
   \left[ {J^{\Dot{a}}}_{\Dot{b}} ,\, {J^{\Dot{c}}}_{\Dot{d}}  \right] &= {\delta^{\Dot{c}}}_{\Dot{b}} {J^{\Dot{a}}}_{\Dot{d}}
   -  {\delta^{\Dot{a}}}_{\Dot{d}} {J^{\Dot{c}}}_{\Dot{b}}\cr
   \left[ {J^{\Dot{a}}}_{\Dot{b}} ,\, J_+^{\Dot{c}\Dot{d}} \right] &= {\delta^{\Dot{c}}}_{\Dot{b}} J_+^{\Dot{a}\Dot{d}}
  + {\delta^{\Dot{d}}}_{\Dot{b}} J_+^{\Dot{c}\Dot{a}} - \frac{1}{3} {\delta^{\Dot{a}}}_{\Dot{b}} J_+^{\Dot{c}\Dot{d}} \cr
   \left[ {J^{\Dot{a}}}_{\Dot{b}} ,\, J^-_{\Dot{c}\Dot{d}} \right] &= -{\delta^{\Dot{a}}}_{\Dot{c}} J^-_{\Dot{b}\Dot{d}}
  - {\delta^{\Dot{a}}}_{\Dot{d}} J^-_{\Dot{c}\Dot{b}} + \frac{1}{3} {\delta^{\Dot{a}}}_{\Dot{b}} J^-_{\Dot{c}\Dot{d}}\cr
   \left[ U ,\, J^-_{\Dot{c}\Dot{d}} \right] &= - J^-_{\Dot{c}\Dot{d}} \qquad
   \left[ U ,\, J_+^{\Dot{c}\Dot{d}} \right] = + J_+^{\Dot{c}\Dot{d}}  \qquad
   \left[ U ,\, {J^{\Dot{c}}}_{\Dot{d}} \right] = 0
\end{aligned}
\end{equation}
The quadratic Casimir of the algebra reads
\begin{equation}
   \mathcal{C}_2\left(\mathfrak{so}\left(10,2\right)\right) =
    \frac{1}{6} {J^{\Dot{a}}}_{\Dot{b}} {J^{\Dot{b}}}_{\Dot{a}} + \frac{1}{9} U^2 + 12 \left( J_+^{\Dot{a}\Dot{b}}
   J^-_{\Dot{a}\Dot{b}} +    J^-_{\Dot{a}\Dot{b}} J_+^{\Dot{a}\Dot{b}}  \right) = I_4 - \frac{99}{16}
\end{equation}
Definition of the grade $\pm 1$ generators goes along the same
lines as for $\mathfrak{e}_{7(-5)}$ so we omit them here. Let us
only note that the quadratic Casimir of the minimal unitary
realization of  $\mathfrak{e}_{7(-25)}$ takes on the same value as
that of $\mathfrak{e}_{7(-5)}$ and equals to $-14$.

\subsection{Truncations to minimal realizations of $SO(2p,2)$
$(p=2,3,4,5)$ as quasiconformal subgroups}

The minimal unitary realization of $\mathrm{E}_{7(-25)}$ can be further
truncated to obtain the minimal unitary realizations of subgroups
of the from $SO(2p,2)$. To this end consider the 5-grading of
$\mathrm{E}_{7(-25)}$
\begin{equation}
    \mathbf{133} = \mathbf{1} \oplus \mathbf{32} \oplus \left( \mathfrak{so}(10,2) \oplus \Delta \right) \oplus \mathbf{32}
    \oplus \mathbf{1}  \nonumber
\end{equation}

The Lie algebra $\mathfrak{so}(10,2)$ has a subalgebra
\begin{equation}
  \mathfrak{so}(2,2) \oplus \mathfrak{so}(8) =
\mathfrak{su}(1,1)_L \oplus \mathfrak{su}(1,1)_R \oplus
\mathfrak{so}(8)
\end{equation}
under which its  spinor representation $\mathbf{32}$
decomposes as\footnote{ In the decomposition of the other spinor
representation $\mathbf{32}'$ the $\mathbf{8}_c$ and $\mathbf{8}_s$ are interchanged. }
\begin{equation}
 \mathbf{32} =
    \left (  \mathbf{2},\mathbf{1},\mathbf{8}_c \right)
       \oplus
     \left(\mathbf{1},\mathbf{2},\mathbf{8}_s \right)
\end{equation}
By restricting ourselves to $\mathrm{SU}(1,1)_R$ singlets we obtain the
minimal unitary realization of $\mathrm{SO}(10,2)$ subgroup of $\mathrm{E}_{7(-25)}$
\begin{equation}
   \mathfrak{so}(10,2) = \mathbf{1} \oplus
                          \left( \mathbf{2},\mathbf{1},\mathbf{8}_c \right)
                         \oplus \left(
                             \mathfrak{su}\left(1,1\right)_L
                              \oplus \mathfrak{so}(8) \oplus \Delta
                           \right)
    \oplus \left(\mathbf{2},\mathbf{1},\mathbf{8}_c \right) \oplus \mathbf{1}
\end{equation}
By restricting to the $\mathrm{SU}(1,1)_L$ singlets one obtains a $\mathrm{SO}(8)$
triality rotated realization of $\mathrm{SO}(10,2)$
\begin{equation}
   \mathfrak{so}(10,2) = \mathbf{1} \oplus \left( \mathbf{1}, \mathbf{2}, \mathbf{8}_s \right)
          \oplus \left( \mathfrak{su}(1,1)_R \oplus \mathfrak{so}(8) \oplus \Delta \right)
          \oplus \left( \mathbf{1}, \mathbf{2}, \mathbf{8}_s \right) \oplus \mathbf{1}
\end{equation}
We recall the decomposition of the three 8 dimensional irreps of
$\mathrm{SO}(8)$ with respect to its $\mathrm{SU}(4) \times \mathrm{U}(1)$ subgroup
\cite{slansky}
\begin{eqnarray}
  \mathbf{8}_v= \mathbf{4}^{(1)} + \bar{\mathbf{4}}^{(-1)} \\
   \mathbf{8}_s= \mathbf{1}^{(2)} + \mathbf{6}^{(0)} + \mathbf{1}^{(-2)} \\
   \mathbf{8}_c= \mathbf{4}^{(-1)} + \bar{\mathbf{4}}^{(1)}
\end{eqnarray}
Thus by splitting the indices $\dot{a}, \dot{b},..$ as
\begin{equation*}
   \dot{a} = (\mu , x) \quad \mu=1,2,3,4 \quad  x=5,8
\end{equation*}
and identifying the generators of the  $\mathrm{U}(4)$ subgroup
of $\mathrm{SO}(8)$
with $J^{\mu}_{\nu} $ and the $\mathrm{SU}(1,1)_L$ generators with $\left( {J^x}_y
-\frac{1}{2}  {\delta^x}_y {J^z}_z \right)$ we find the following decompositions of the
oscillators with respect to the $\mathrm{SU}(1,1)_L \times \mathrm{SU}(1,1)_R \times
\mathrm{SO}(8)$
\begin{equation}
\begin{split}
  \left( \tilde{Z}_{\mu x}, Z^{\nu , y} \right) = \left(\mathbf{2},\mathbf{1},\mathbf{8}_c \right) \cr
   \left( \tilde{Z}_{\mu \nu}, Z^{\mu \nu}, \tilde{Z}_{xy}, Z^{xy}, \tilde{Z}_{67} , Z^{67} \right) =
   \left( \mathbf{1}, \mathbf{2}, \mathbf{8}_s \right)
\end{split}
\end{equation}
By setting either set of these 16 operators equal to zero we get a
consistent truncation of the minimal unitary realization of
$\mathrm{E}_{7(-25)}$ to one of its $\mathrm{SO}(10,2)$ quasiconformal subgroups.
Here we give the realization obtained by setting the operators in
the $( \mathbf{1}, \mathbf{2}, \mathbf{8}_s)$ representation equal to zero.

The $\mathfrak{so}(8)$ generators in the grade zero subspace of
$\mathfrak{so}(10,2)$ are given by
\begin{equation*}
  \mathfrak{so}(8) = ( J^-_{\mu\nu}, {J^{\mu}}_{\nu}, J_+^{\mu\nu} )
\end{equation*}
where
\begin{equation}
{J^{\mu}}_{\nu}= 2 Z^{\mu x}\tilde{Z}_{\nu x} - \frac{1}{2}
{\delta^{\mu}}_{\nu} ( Z^{\rho z} \tilde{Z}_{\rho z} )
\qquad
\begin{aligned}
J^-_{\mu\nu} = -\frac{1}{6} \epsilon_{\mu\nu\rho\lambda} Z^{\rho
x}Z^{\lambda y} \epsilon_{xy} \cr
J_+^{\mu\nu} = -\frac{1}{6} \epsilon^{\mu\nu\rho\lambda}
\tilde{Z}_{\rho x} \tilde{Z}_{\lambda y} \epsilon^{xy}
\end{aligned}
\end{equation}
The quadratic Casimir of the grade zero subalgebra
$\mathfrak{so}(8) \oplus \mathfrak{su}(1,1)$ takes the form
\begin{equation}
 C_2(\mathfrak{so}(8) \oplus \mathfrak{su}(1,1)) = \frac{1}{6} (
{J^{\mu}}_{\nu} {J^{\nu}}_{\mu} +{J^x}_y {J^y}_x + U^2 ) + 12
( J_+^{\mu\nu} J^-_{\mu\nu} + J^-_{\mu\nu} J_+^{\mu\nu} )
\end{equation}
where
\begin{equation}
  {J^x}_y = 2Z^{\mu x} \tilde{Z}_{\mu y} - \delta^x_y
(Z^{\mu z} \tilde{Z}_{\mu z} )
\end{equation}
and
\begin{equation}
 U=-\frac{1}{2} \left( Z^{\mu x} \tilde{Z}_{\mu x} + \tilde{Z}_{\mu x}
 Z^{\mu x} \right)
\end{equation}
The grade -1 generators of $\mathrm{SO}(10,2)$ are simply given by
\begin{equation}
 E^{\mu x} =y Z^{\mu x} \quad \tilde{E}_{\mu x} = y \tilde{Z}_{\mu x}
\end{equation}
and those of grade +1 are given by
\begin{equation}
  F^{\mu x} = - p Z^{\mu x} + \frac{2\, i}{y} \left[ Z^{\mu x}, I_4 \right]
  \qquad
  \Tilde{F}_{\mu x} = - p \Tilde{Z}_{\mu x} + \frac{2\, i}{y} \left[ \Tilde{Z}_{\mu x}, I_4 \right]
\end{equation}
where grade +2 generator $F$ is given by
\begin{equation}
 F= \frac{1}{2} p^2 + \frac{2}{y^2} I_4 \qquad
  I_4 = \mathcal{C}_2\left( \mathfrak{so}(8) \oplus \mathfrak{su}(1,1) \right) +  \frac{73}{48}
\end{equation}
The algebra's quadratic Casimir equals
\begin{equation}
   \mathcal{C}_2 \left( \mathfrak{so}\left(10,2\right)\right) = -4
\end{equation}

To obtain the truncations of the above realization of $\mathfrak{so}(10,2)$ to the minimal unitary realizations
of $\mathfrak{so}(2p+2,2)$ we need only restrict the indices $\mu, \nu,...$ of non-vanishing oscillators to run
over $\mu, \nu,.. =1,..,p$, where $p=1,2,3$.

\section{Discussion}

We note that the quadratic Casimir of quasi-conformal algebras evaluated on the minimal
realization is related to the algebra's dual Coxeter number $g^\vee$
\begin{equation}
    \mathcal{C}_2 = - \frac{1}{108} g^\vee \left( 5 g^\vee - 6 \right)
\end{equation}
for cases where $g^0$ is simple. 
This is a reflection of the fact that quasi-conformal algebras can
be constructed in a unified manner. In a forthcoming paper we will
give such a unified approach to the construction of the minimal
unitary representations of all simple noncompact groups
\cite{gpfuture}.

Here we would like to stress that the minimal unitary realizations given above, in \cite{GKN:2} as well as in the
unified approach \cite{gpfuture} correspond to quantization of the geometric action of the respective noncompact
group as a quasiconformal group as defined and studied for the exceptional groups in \cite{GKN:1}. A
quasiconformal group $G$  leaves invariant a generalized light-cone with respect to a distance function defined
in terms of the quartic invariant of its subgroup $H$ which is the normalizer of the $\mathrm{SL}(2,\mathbb{R})$ subgroup
generated by grade $\pm 2$ elements  of its Lie algebra $\mathfrak{g}$ \footnote{ For some of the classical
noncompact groups the corresponding quartic invariant may be degenerate \cite{gpfuture}. }. The realization of
this $\mathrm{SL}(2,\mathbb{R})$ subgroup inside the minimal unitary realization is precisely of the form that arises in
conformal quantum mechanics \cite{fubini} as was stressed in \cite{GKN:2}. The quartic invariant of the subgroup
$H$ plays the role of coupling constant in the corresponding conformal quantum mechanics.


\section*{Appendix}
\appendix
\section{Going from $\mathfrak{su}^\ast(8)$ to $\mathfrak{su}(6,2)$ basis}

Recall that position and momentum operators $X^{AB}$ and $P_{AB}$ transform as $\mathbf{28}$ and
$\Tilde{\mathbf{28}}$ under $\mathfrak{su}^\ast(8)$. To build annihilation and creation operators we need to take
complex linear combinations  of the form $X^{AB} \pm i P_{AB}$, which transform covariantly under
$\mathfrak{so}^\ast(8)$ subalgebra of $\mathfrak{su}^\ast(8)$.  We expect resulting creation and annihilation
operators to transform as $\mathbf{28}$ and $\overline{\mathbf{28}}$ of some non-compact form of
$\mathfrak{su}(8)$ \footnote{ Notice that compact $\mathfrak{su}(8)$ is not a subalgebra of
$\mathfrak{e}_{7(-25)}$.}. The isomorphism $\mathfrak{so}^\ast(8) \simeq \mathfrak{so}(6,2)$ suggests that this
non-compact form should be $\mathfrak{su}(6,2)$ as we shall establish.

In order to elucidate the role of triality of $\mathfrak{so}(8)$ we recall that
adjoint representation of compact $\mathfrak{e}_7$ decomposes into four representations
of $\mathfrak{so}(8)$:
\begin{equation*}
   \mathbf{133} = \mathbf{28} \oplus \mathbf{35}_v\oplus \mathbf{35}_s \oplus \mathbf{35}_c
\end{equation*}
where three $\mathbf{35}$ correspond to symmetric traceless tensor in $\mathbf{8}_v \otimes \mathbf{8}_v$,
$\mathbf{8}_s \otimes \mathbf{8}_s$ and $\mathbf{8}_c \otimes \mathbf{8}_c$ respectively, with $\mathbf{8}_v$,
$\mathbf{8}_s$  and $\mathbf{8}_c$ being three inequivalent eight dimensional representations of
$\mathfrak{so}(8)$. Triality of $\mathfrak{so}(8)$ then maps $\mathbf{35}$ representations into one another.
Observe also, that $\mathbf{28}$ combined with any one of three $\mathbf{35}$ generate an $\mathfrak{su}(8)$
subalgebra of $\mathfrak{e}_7$. Compact  $\mathfrak{so}(8)$ becomes $\mathfrak{so}^\ast\left(8\right)$ if we
consider $\mathfrak{e}_{7(-25)}$ instead of compact $\mathfrak{e}_7$ and $\mathfrak{su}(8)$ becomes
$\mathfrak{su}^\ast(8)$.

Consider the Clifford algebra of $\mathbb{R}^{6,2}$
\begin{equation}
   \left\{ \Gamma^a \,, \Gamma^b \right\} = 2 \eta^{ab}
\end{equation}
and  choose a basis with the following hermiticity property
\begin{equation}
    \left( \Gamma^{a} \right)^\dagger = \eta_{ab} \Gamma^{b} = \omega \cdot \Gamma^{a} \cdot \omega^{-1}
\end{equation}
where $\omega = \Gamma^7 \cdot \Gamma^8$ is a $16 \times 16$ symplectic matrix.
One particular choice of basis, in which chirality matrix $\Gamma^9$ is diagonal, is given as follows
\begin{equation}
\begin{aligned}
   &\Gamma^1 = \block{\sigma_1}{\mathbb{I}_2}{\mathbb{I}_2}{\mathbb{I}_2} \qquad
   &\Gamma^2 = \block{\sigma_2}{\sigma_1}{\mathbb{I}_2}{\sigma_2} \\
   &\Gamma^3 = \block{\sigma_2}{\sigma_2}{\sigma_2}{\sigma_2} \qquad
   &\Gamma^4 = \block{\sigma_2}{\sigma_2}{\sigma_3}{\mathbb{I}_2} \\
   &\Gamma^5 = \block{\sigma_2}{\sigma_3}{\mathbb{I}_2}{\sigma_2} \qquad
   &\Gamma^6 = \block{\sigma_2}{\sigma_2}{\sigma_1}{\mathbb{I}_2}\\
   &\Gamma^7 = i \block{\sigma_2}{\mathbb{I}_2}{\sigma_2}{\sigma_3} \qquad
   &\Gamma^8 = i \block{\sigma_2}{\mathbb{I}_2}{\sigma_2}{\sigma_1}
\end{aligned}
\end{equation}
Then,
\begin{equation}
\begin{split}
     Z^{ab} &= \frac{1}{4} \, {\Gamma^{ab}}_{CD}  \left( X^{CD} - i P_{CD} \right) \\
     \Tilde{Z}^{ab} &= \frac{1}{4} \, {\Gamma^{ab}}_{CD}  \left( X^{CD} + i P_{CD} \right)
\end{split}
\end{equation}
where transformation coefficient are given by matrix elements of chiral representation of
$\mathfrak{so}(6,2)$ generators
\begin{equation}
   {\Gamma^{ab}}_{CD} = \mathbb{P}\left(\frac{i}{4} \left[ \Gamma^{a} \,, \Gamma^{b} \right]\right)_{CD}
\end{equation}
and $\mathbb{P}$ is the chiral projection operator in spinor space.
Symplectic structure \eqref{eq:XPsympl} of $X$ and $P$ induces the symplectic structure
\begin{equation}
  \left[ \Tilde{Z}^{ab} \,, Z^{cd} \right] = \frac{1}{8} \mathop{\text{Tr}} \left[ \Gamma^{ab} \, \Gamma^{cd} \right]
      = \frac{1}{2} \left( \eta^{ca}\eta^{db} - \eta^{cb}\eta^{da} \right) \,.
  \label{eq:ZZsympl}
\end{equation}
on $Z$ and $\Tilde{Z}$. Gamma matrices defined above satisfy the following identities
\begin{gather*}
  {\Gamma^{ab}}_{AB} = - {\Gamma^{ab}}_{BA} = - {\Gamma^{ba}}_{AB} \\
  {\Gamma^{abcd}}_{AB} = \frac{1}{24} {\epsilon^{abcd}}_{efgh} {\Gamma^{efgh}}_{AB} = {\Gamma^{abcd}}_{BA} \\
  {\Gamma^{abcd}}_{ABCD} :=  {\Gamma^{[ab}}_{[AB} {\Gamma^{cd]}}_{CD]}  \\
  {\Gamma^{abcd}}_{ABCD} =
             - \frac{1}{24} {\epsilon^{abcd}}_{efgh} {\Gamma^{efgh}}_{ABCD}
             = - \frac{1}{24} \epsilon_{ABCDEFGH} {\Gamma^{abcd}}_{EFGH}  \\
  {\Gamma^{ac}}_{[AB} {\Gamma^{cb}}_{CD]} = {\Gamma^{bc}}_{[AB} {\Gamma^{ca}}_{CD]}
    = \frac{1}{24} \epsilon_{ABCDEFGH} {\Gamma^{ac}}_{[EF} {\Gamma^{cb}}_{GH]}
\end{gather*}
where
\begin{equation*}
   {\Gamma^{abcd}}_{AB} = \mathbb{P} \left( \Gamma^{[a} \Gamma^b \Gamma^c \Gamma^{d]} \right)_{AB}  \,.
\end{equation*}
These identities allow us to rewrite generators of $\mathfrak{e}_{7(-25)}$ in $\mathfrak{su}(6,2)$ basis:
\begin{equation}
\begin{aligned}
   \eta^{bc} {J^a}_{c} - \eta^{ac} {J^b}_{c} &=  {\Gamma^{ab}}_{AB} \left( {J^A}_B -  {J^B}_A \right) \\
   \eta^{bc} {J^a}_{c} + \eta^{ac} {J^b}_{c} &= {\Gamma^{ab}}_{ABCD} \left( J^{ABCD} +
                                           \left(\epsilon J \right)_{ABCD} \right) \\
   J^{abcd} + \frac{1}{24} {\epsilon^{abcd}}_{efgh} J^{efgh} &=
                         {\Gamma^{abcd}}_{AB} \left( {J^A}_B + {J^B}_A \right) \\
   J^{abcd} - \frac{1}{24} {\epsilon^{abcd}}_{efgh} J^{efgh} &= {\Gamma^{abcd}}_{ABCD}
                                        \left( J^{ABCD} - \left(\epsilon J \right)_{ABCD} \right)
\end{aligned}
\end{equation}
or, more succintly,
\begin{equation}
\begin{split}
   \eta^{bc} {J^{a}}_c &=  {\Gamma^{ab}}_{AB} {J^{A}}_B + {\Gamma^{ab}}_{ABCD} J^{ABCD} \\
   J^{abcd} &= {\Gamma^{abcd}}_{AB} {J^{A}}_B +  {\Gamma^{abcd}}_{ABCD} J^{ABCD}
\end{split}
\end{equation}

\section{Minimal realization of $\mathfrak{e}_{8(8)}$ in $\mathfrak{su}^\ast(8)$ basis}

Non-compact exceptional Lie algebra $\mathfrak{e}_{8(8)}$ also admits realization in an
$\mathfrak{su}^\ast(8)$ basis. It is seen via the following chain of subalgebra inclusions
$\mathfrak{su}^\ast(8) \subset \mathfrak{e}_{7(7)} \subset \mathfrak{e}_{8(8)}$.

Algebra $\mathfrak{e}_{7(7)}$ is generated as
\begin{equation} \label{eq:e7(7)generators}
\begin{split}
  {J^A}_B &= -2 i X^{AC} P_{CB} -  \frac{i}{4} {\delta^A}_B  X^{CD}  P_{CD} \\
  J^{ABCD} &= - \frac{i}{2} X^{[AB} X^{CD]} + \frac{i}{48} \epsilon^{ABCDEFGH} P_{EF} P_{GH} \,.
\end{split}
\end{equation}
where $A, B, \dots$ are $\mathfrak{su}^\ast(8)$ indices.
Note different relative signs between $XX$ and $PP$ terms in \eqref{eq:e7(7)generators} and
\eqref{eq:e7-25generators}. It amounts to change of sign in the commutator on the third line
\begin{equation}
\begin{split}
  \left[ {J^A}_B, {J^C}_D \right] & = {\delta^C}_B {J^A}_D - {\delta^A}_D {J^C}_B \\
  \left[ {J^A}_B, J^{CDEF} \right] &= - 4 {\delta^{[C}}_B J^{DEF]A} - \frac{1}{2} {\delta^A}_B J^{CDEF} \\
  \left[ J^{ABCD}, J^{EFGH} \right] &= + \frac{1}{36} \epsilon^{ABCDK[EFG} {J^{H]}}_K
\end{split}
\label{eq:e7(7)}
\end{equation}
as compared to that in \eqref{eq:e7} while does not change the hermiticity properties \eqref{eq:realityConditions}
resulting in the following quadratic Casimir
\begin{equation}
\begin{split}
 \mathcal{C}_2 & =  \frac{1}{6} {J^A}_B {J^B}_A + \frac{1}{24} \epsilon_{ABCDEFGH} J^{ABCD} J^{EFGH} \\
               & =  \frac{1}{6} {J^A}_B {J^B}_A + J^{ABCD} (\epsilon J)_{ABCD} \,.
\end{split}
 \label{eq:e7(7)C2}
\end{equation}
The decomposition of $\mathfrak{e}_{7(7)}$ with respect to the maximal compact subalgebra $\mathfrak{usp}(8)$
of $\mathfrak{su}^\ast(8)$ results now in
\begin{equation*}
  \mathbf{133} =\mathbf{63} \oplus \mathbf{70} = \left( \mathbf{36}_\text{c.} \oplus \mathbf{27}_\text{n.c.}  \right)
    \oplus \left( \mathbf{42}_\text{n.c.} \oplus \mathbf{27}_\text{c.} \oplus \mathbf{1}_\text{n.c.} \right)
\end{equation*}
and shows the the constructed $\mathfrak{e}_7$ is indeed $\mathfrak{e}_{7(7)}$. The remaining
generators of algebra $\mathfrak{e}_{8(8)}$ are then given by
\begin{equation*}
  E^{AB} = -i y X^{AB} \qquad
  \Tilde{E}_{AB} = -i y P_{AB} \qquad E= -\frac{i}{2} y^2
\end{equation*}
and
\begin{equation*}
\begin{aligned}
  &F = \frac{1}{2 i} p^2 + \frac{2}{i y^2} I_4 \left( X\,, P \right)\\
  &I_4 \left( X\,, P \right) = \mathcal{C}_2 + \frac{323}{16}
\end{aligned}
 \qquad
\begin{aligned}
   F^{AB} &= i p X^{AB} + \frac{2}{y} \left[ X^{AB} \,, I_4 \left( X, P \right) \right] \\
   \Tilde{F}_{AB} &= i p P_{AB} + \frac{2}{y} \left[ P_{AB} \,, I_4 \left( X, P \right) \right] .
\end{aligned}
\end{equation*}
They satisfy the same commutation relations as their counterparts of $\mathfrak{e}_{8(-24)}$ except
for
\begin{equation}\begin{split}
   \left[ J^{ABCD} \,, E^{EF} \right] &= + \frac{1}{24} \epsilon^{ABCDEFGH} \Tilde{E}_{GH} \, \\
   \left[ J^{ABCD} \,, F^{EF} \right] &= + \frac{1}{24} \epsilon^{ABCDEFGH} \Tilde{F}_{GH} \, \\
   \left[ \Tilde{E}_{AB} \,, \Tilde{F}_{CD} \right] &= + 12\, {(\epsilon J)}_{ABCD}
\end{split}\end{equation}

\end{document}